\documentclass[10pt]{iopart}

\usepackage{graphicx}
\usepackage{amssymb}
\usepackage{exscale}
\usepackage{iopams}
\usepackage{subfigure}
\usepackage{xspace}
\usepackage{citesort}

\renewcommand{\Im}[0]{\mathrm{Im}\,}

\newcommand{\ie}[0]{i.e.\@\xspace}
\newcommand{\eg}[0]{e.g.\@\xspace}

\newcommand{\om}[0]{\omega}
\newcommand{\gammab}[0]{\overline{\gamma}}
\newcommand{\Ep}{E_\mathrm{p}}
\newcommand{\omb}[0]{\overline{\omega}}

\newcommand{\Ap}[0]{A_\mathrm{p}}
\newcommand{\Ae}[0]{A_\mathrm{e}}
\newcommand{\nag}{\phantom{\dag}}

\newcommand{\las}[0]{\langle}
\newcommand{\ras}[0]{\rangle}

\renewcommand{\tilde}[1]{\widetilde{#1}}

\begin{document}

\title{Spectral functions of the spinless Holstein model}

\author{J Loos\dag, M Hohenadler\ddag, and H Fehske\S}

\address{\dag\ %
  Institute of Physics, Academy of Sciences of the Czech Republic, Prague
}

\address{\ddag\ %
  Institute for Theoretical and Computational Physics, TU Graz, Austria
}

\address{\S\ %
  Institute for Physics, Ernst-Moritz-Arndt University Greifswald, Germany
}

\ead{\mailto{hohenadler@itp.tugraz.at}}

\begin{abstract}
  An analytical approach to the one-dimensional spinless Holstein model is
  proposed, which is valid at finite charge-carrier concentrations.  Spectral
  functions of charge carriers are computed on the basis of self-energy
  calculations. A generalization of the Lang-Firsov canonical transformation
  method is shown to provide an interpolation scheme between the extreme
  weak-- and strong-coupling cases.  The transformation depends on a
  variationally determined parameter that characterizes the charge
  distribution across the polaron volume. The relation between the spectral
  functions of polarons and electrons, the latter corresponding to the
  photoemission spectrum, is derived.  Particular attention is paid to the
  distinction between the coherent and incoherent parts of the spectra, and
  their evolution as a function of band filling and model parameters. Results are
  discussed and compared with recent numerical calculations for the
  many-polaron problem.
\end{abstract}


\section{Introduction}\label{sec:introduction}

Experiments on a variety of novel materials, ranging from
quasi-one-dimensional (1D) MX solids \cite{mx}, organics \cite{org} and
quasi-2D high-$T_{\rm c}$ cuprates \cite{htc} to 3D colossal-magnetoresistive
manganites~\cite{mang,edwards}, provide clear evidence for the existence of
polaronic carriers, \ie, quasiparticles consisting of an electron and a
surrounding lattice distortion.  This has motivated considerable theoretical
efforts to achieve a better understanding of strongly coupled electron-phonon
systems in the framework of microscopic models.

Unfortunately, even for highly simplified models, such as the spinless
Holstein model \cite{Ho57} considered here, no exact analytical solutions
exist, except for the Holstein polaron problem with a relativistic dispersion
\cite{MeScGu94} or in infinite dimensions \cite{Su72}.  As a consequence,
numerous numerical studies have been carried out, focussing either on the
empty band limit (\ie, one or two electrons only; see
\cite{RT92,WRF96,KTB02,HoEvvdL03,Spencer} and references therein), or on the
half-filled band case in one dimension, where the Peierls transition takes
place \cite{BuMKHa98}. In contrast, very little work has been done at finite
carrier densities away from half filling \cite{HuZheng,Datta}, which are,
however, often realized
in experiment \cite{mx,org,htc,mang,edwards}. Recently, this so-called {\it
  many-polaron problem} has been addressed numerically
\cite{CaGrSt99,HoNevdLWeLoFe04,HoWeAlFe05}. The results have led to a fairly
good understanding of many aspects, but their interpretation is not always
straight forward, which makes analytical calculations along these lines
highly desirable.

In this paper, we propose an analytical approach to the 1D spinless Holstein
model, capable of describing finite charge-carrier concentrations at
arbitrary model parameters, including the important adiabatic
intermediate-coupling (IC) regime. 

First, using standard perturbation theory based on the
self-energy calculation, the spectral functions of charge carriers will be
determined in the weak-coupling (WC) and the strong-coupling (SC) limits at
zero temperature ($T=0$). In the SC regime, the relation between the spectral
function of polarons, determining the equilibrium properties (in particular,
the chemical potential), and the electronic spectral function, determining
the photoemission spectrum, will be discussed. Special emphasis will be laid
on the distinction between the coherent and the incoherent parts of the
spectra, which may be calculated separately within the present
approach. 

Furthermore, using a generalization
of the Lang-Firsov canonical transformation method~\cite{LangFirsov}, an
interpolation scheme between the extreme WC and SC cases will be proposed. In
particular, the canonical transformation will depend on the distance $R$
characterizing the charge distribution across the polaron volume. For a given
set of model parameters and carrier concentration $n$, $R$ will be determined
from the minimum of the total energy given by the transformed Hamiltonian in
the first, Hartree-like approximation. With $R$ found in this way, the
polaronic and electronic spectral functions will be calculated to study
their dependence on the model parameters and the carrier density. The results
will be discussed with regard to recent numerical calculations
\cite{HoNevdLWeLoFe04,HoWeAlFe05}, which have revealed a cross-over from a
system with polaronic carriers to a rather metallic system with increasing
band filling $n$ in the intermediate electron-phonon coupling regime.

The paper is organized as follows. In section~\ref{sec:model} we introduce the
model used, whereas in section~\ref{sec:GFapproach} we
derive the analytical results for WC, SC and IC regimes.  Our results are
discussed in section~\ref{sec:results}, and section~\ref{sec:conclusion}
contains our conclusions.

\section{Theory}\label{sec:theory}

\subsection{Model}\label{sec:model}
 
In this paper, we are exclusively concerned with the Holstein model (HM) of
spinless fermions, which describes electrons coupled locally to Einstein
phonons. Although we shall use different canonical transformations to
describe the IC and SC regime later on, the Hamiltonian can be written in the general
form
\begin{equation}\label{eq:HM}
  H 
  =
    \eta\sum_i c^\dag_i c^{\nag}_i 
  - \sum_{i,j}C_{ij}^{\nag} c_{i}^{\dag} c_{j}^{\nag}
  + \om_0\sum_i  
  ( b_i^{\dag} b_i^{\nag} + \mbox{\small $\frac{1}{2}$})
  \,,
\end{equation}
where the definition of $\eta$ and $C_{ij}$ will be different depending on
the approach used. Here $c_{i}^{\dag}$ ($c_{i}^{\nag}$) creates (annihilates)
a spinless fermion at site $i$, $b_i^{\dag}$ and $b_i^{\nag}$ are bosonic
operators for the dispersionless phonons of energy $\om_0$ ($\hbar=1$), and the strength of the
electron-phonon interaction is specified by the dimensionless coupling
constants $\lambda=\Ep/2t$ and $g=\sqrt{\Ep/\om_0}$ in the adiabatic
($\om_0/t\ll1$) and anti-adiabatic ($\om_0/t\gg1$) regimes, respectively,
where $\Ep$ is the well-known polaron binding energy in the atomic limit
[$C_{ij}=0$ for $i\neq j$ in equation~(\ref{eq:HM})].

In the WC case, in which we use the original, untransformed Holstein
Hamiltonian, we have $\eta=-\mu$, where $\mu$ denotes the chemical
potential, and non-zero coefficients
\begin{equation} \label{eq:wcC}
  C_{ii} 
  =
  g \om_0  ( b_i^{\dag} + b_i^{\nag}) 
  \,,\qquad 
  C_{\las ij\ras}
  =
  t
  \,.
\end{equation}
In contrast, the starting point in the SC regime will be the Hamiltonian with
$\eta=-g^2\om_0-\mu=-\Ep-\mu$ and
\begin{equation}\label{eq:scC}
  C_{ii}
  =
  0 
  \,,\qquad 
  C_{\las ij\ras}
  =
  t \rme^{-g( b_i^{\dag} - b_i^{\nag} - b_j^{\dag} + b_j^{\nag})} 
  \,.
\end{equation}
%

\subsection{Green functions approach}\label{sec:GFapproach}

We treat the HM~(\ref{eq:HM}) using the formalism of the generalized
Matsubara Green functions introduced by Kadanoff and Baym \cite{KaBa62} and
Bonch-Bruevich and Tyablikov \cite{BBTy62}, and applied to the single-polaron
problem by Schnakenberg \cite{Sc66}. The Green function equation of motion
deduced from $H$ may be converted into an equation for the self-energy of the
spinless fermions, and can be solved by iteration (see
\cite{Sc66,Lo94,FeLoWe97} for details). In the second iteration step, the
self-energy is obtained in the form
\begin{eqnarray}\label{eq:SEA}
  \fl
  \Sigma(j_1 \tau_1;j_2\tau_2)
  =
  -\las C_{j_1j_2}\ras 
  \delta(\tau_1-\tau_2)
  \nonumber\\
  +\sum_{j' j''}
  G^\mathrm{M}(j' \tau_1;j''\tau_2)
  [\las T_{\tau} C_{j_1 j'}
  (\tau_1)C_{j'' j_2}(\tau_2)
  \ras -\las C_{j_1 j'}\ras
  \las C_{j'' j_2}\ras]
  \,,
\end{eqnarray}
where $G^\mathrm{M}(j_1\tau_1;j_2\tau_2)$ represents the first-order
fermionic Matsubara Green function, and the symbol $T_{\tau}$ denotes the
time ordering operator acting on the imaginary times $\tau_i$.
Fourier-transforming both sides of equation~(\ref{eq:SEA}), carrying out the
standard summation over the Matsubara boson frequencies \cite{Ma90}, and
using the analytical continuation in the complex frequency plane, the
retarded momentum-- and energy-dependent fermion Green function follows as
\begin{equation}
  G^\mathrm{R}(k,\omb)
  =
  \frac{1}{\omb - (\xi_k+\eta)-\Sigma(k,\omb)}
  \,,
\end{equation}
where $\xi_k$ is the fermionic band dispersion in the first approximation,
$\omb=\om+\rmi\epsilon$ ($\epsilon\rightarrow 0^+$), and $\Sigma(k,\om)$
denotes the Fourier transform of the collisional part of the
self-energy given by the second term on the r.~h.~s of
equation~(\ref{eq:SEA}). The related, normalized spectral function is given by
\begin{equation}
  A(k,\om)
  =
  -\frac{1}{\pi}\mathrm{Im}\,G^\mathrm{R}(k,\om+\rmi 0^+)
  \,.
\end{equation}

Although the insulating Peierls phase with long-range charge-density-wave
order, which is the ground state of the half-filled spinless Holstein model
above a critical coupling strength depending on $\om_0$ \cite{BuMKHa98}, can
in principle be incorporated into the present theory, this has not been done
here, as we are interested in intermediate band fillings away from $n=0.5$.
Furthermore, we neglect any extended pairing for $0<n<0.5$, as well as
the possible formation of a polaronic superlattice in the half-filled band
case ($n=0.5$). Finally, the possibility of phase separation, which can in principle occur in
the present model, is not considered.

\subsubsection{Weak coupling}\label{sec:weak-coupling}

In the WC limit the self-energy is given by
\begin{equation}\label{eq:wc_Sigma}
  \fl
  \Sigma(k,\omb)
  =
  \frac{\om_0\Ep}{\pi W}\int_{-W}^W
  \frac{\rmd\xi}{\sqrt{1-(\xi/W)^2}}
  \left[
    \frac{1-n_\mathrm{F}(\xi - \mu)}{\omb - \om_0 - (\xi -
      \mu)}
    +
    \frac{n_\mathrm{F}(\xi - \mu)}{\omb + \om_0 - (\xi -
      \mu)}
  \right]
  \,,
\end{equation}
with the Fermi function $ n_\mathrm{F}(\om)=(\rme^{\beta \om}+1)^{-1}$ and
the bare bandwidth $W=2t$. At $T=0$, and defining $\xi_0=\max{(\mu,-W)}$, we get
\begin{eqnarray}\label{eq:wc_re_Sigma}
  \fl
  \mathrm{Re}\,\Sigma(k,\om)
  =
  \frac{\om_0\Ep}{\pi W}
  \left[
    \mathcal{P}\int_{-W}^{\xi_0}
    \frac{\rmd\xi}{\sqrt{1-(\xi/W)^2}}
    \frac{1}{\om + \om_0 - (\xi-\mu)}
  \right.\nonumber\\
  +
  \left.
    \mathcal{P}\int_{\xi_0}^W
    \frac{\rmd\xi}{\sqrt{1-(\xi/W)^2}}
    \frac{1}{\om - \om_0 - (\xi-\mu)}
  \right]
\end{eqnarray}
and
\begin{eqnarray}\label{eq:im_Sigma}
  \fl
  \mathrm{Im}\,\Sigma(k,\om)
  =
  -\frac{\om_0\Ep}{W}
  \left[
    \int_{-W}^{\xi_0}
    \frac{\rmd\xi}{\sqrt{1-(\xi/W)^2}}
    \delta[\om + \om_0 - (\xi-\mu)]
  \right.\nonumber\\
  +
  \left.
    \int_{\xi_0}^W
    \frac{\rmd\xi}{\sqrt{1-(\xi/W)^2}}
    \delta[\om - \om_0 - (\xi-\mu)]
  \right]
  \,.
\end{eqnarray}

In the sequel, we shall distinguish between coherent and incoherent
contributions to the single-particle spectral functions, as defined by a zero and non-zero
imaginary part of the self-energy, respectively.
The coherent part of the spectrum is given by 
\begin{equation}\label{eq:wc_A_coh}
  A^\mathrm{c}(k,\om)
  =
  z_k \delta[\om - (E_k - \mu)]
  \,,
\end{equation}
where the renormalized band energy is the solution of
\begin{eqnarray}\label{eq:wc_Ek}
  \fl
  E_k 
  =
  \xi_k +
  \frac{\om_0\Ep}{\pi W}
  \left[
    \mathcal{P}\int_{-W}^{\xi_0}
    \frac{\rmd\xi}{\sqrt{1-(\xi/W)^2}}
    \frac{1}{E_k + \om_0 - \xi}
  \right.\nonumber\\
  +
  \left.
    \mathcal{P}\int_{\xi_0}^W
    \frac{\rmd\xi}{\sqrt{1-(\xi/W)^2}}
    \frac{1}{E_k - \om_0 - \xi}
  \right]
  \,,
\end{eqnarray}
with $\xi_k = - W\cos{k}$ (the bare band dispersion),
and the spectral weight takes the form
\begin{eqnarray}\label{eq:wc_zk}
  \fl
  z_k^{-1}
  =
  \bigg|
    1 +
    \frac{\om_0\Ep}{\pi W}
    \left[
      \mathcal{P}\int_{-W}^{\xi_0}
      \frac{\rmd\xi}{\sqrt{1-(\xi/W)^2}}
      \frac{1}{(E_k + \om_0 - \xi)^2}
      \right.
      \nonumber\\
      +\left.
      \mathcal{P}\int_{\xi_0}^W
      \frac{\rmd\xi}{\sqrt{1-(\xi/W)^2}}
      \frac{1}{(E_k - \om_0 - \xi)^2}
    \right]
  \bigg|
  \,.
\end{eqnarray}

For the incoherent part of the spectral function we find
\begin{equation}\label{eq:wc_A_inc1}
  \fl
  A^{\mathrm{ic}}(k,\om<-\om_0)
  =
  \frac{1}{\pi}
  \frac{\om_0\Ep\left[W^2-(\om+\om_0+\mu)^2\right]^{\frac{1}{2}}\int_{-W}^{\xi_0}\rmd\xi\delta(\om+\om_0+\mu-\xi)}
  {\left[W^2-(\om+\om_0+\mu)^2\right]\left(\om+\mu-\xi_k-\mathrm{Re}\,\Sigma(k,\om)\right)^2 + (\om_0\Ep)^2}
\end{equation}
and
\begin{equation}\label{eq:wc_A_inc2}
  \fl
  A^{\mathrm{ic}}(k,\om>\om_0)
  =
  \frac{1}{\pi}
  \frac{\om_0\Ep\left[W^2-(\om-\om_0+\mu)^2\right]^{\frac{1}{2}}\int_{\xi_0}^W\rmd\xi\delta(\om-\om_0+\mu-\xi)}
  {\left[W^2-(\om-\om_0+\mu)^2\right]\left(\om+\mu-\xi_k-\mathrm{Re}\,\Sigma(k,\om)\right)^2 + (\om_0\Ep)^2}
  \,.
\end{equation}

Finally, the chemical potential $\mu$ for a given electron density $n$ is
determined by
\begin{equation}\label{eq:wc_mu_n}
  \frac{1}{N}
  \sum_k\int_{-\infty}^\infty \rmd \om A(k,\om)
  n_\mathrm{F}(\om)
  =
  n
  \,.
\end{equation}
%

\subsubsection{Strong coupling}\label{sec:strong-coupling}

Hamiltonian~(\ref{eq:HM}) with the coefficients~(\ref{eq:scC}) represents the
Hamiltonian of small polarons, which are the correct quasiparticles in the
SC limit. Using the procedure outlined above, we obtain the polaron
self-energy as  
\begin{eqnarray}
  \fl
  \Sigma(k,\omb)
  =
  \frac{\tilde{W}}{2\pi}\sum_{s\geq1}
  \int_{-\tilde{W}}^{\tilde{W}}
  \frac{\rmd\xi f(k,\xi,s)}{\sqrt{1-(\xi/\tilde{W})^2}}
  \left[
    \frac{1-n_{\mathrm{F}}(\xi+\eta)}{\omb - s\om_0-(\xi+\eta)}
    +
    \frac{n_{\mathrm{F}}(\xi+\eta)}{\omb + s\om_0-(\xi+\eta)}
  \right]\,.
\end{eqnarray}
At $T=0$, we have $\tilde{W}=W \rme^{-g^2}$,
\begin{eqnarray}
  \fl
  \mathrm{Re}\,\Sigma(k,\om)
  =
  \frac{\tilde{W}}{2\pi}\sum_{s\geq1}
  \mathcal{P}\int_{-\tilde{W}}^{\tilde{W}}\frac{\rmd\xi}{\sqrt{1-(\xi/\tilde{W})^2}}
  f(k,\xi,s)
  \nonumber\\
  \times
  \left[
   \frac{\theta(\xi+\eta)}{\om-s\om_0-(\xi+\eta)}
    +
   \frac{\theta(-\xi-\eta)}{\om+s\om_0-(\xi+\eta)}
  \right]
\end{eqnarray}
and
\begin{eqnarray}
  \fl
  \mathrm{Im}\,\Sigma(k,\om)
  =
  -\frac{\tilde{W}}{2}\sum_{s\geq1}
  \int_{-\tilde{W}}^{\tilde{W}}\frac{\rmd\xi}{\sqrt{1-(\xi/\tilde{W})^2}}
  f(k,\xi,s)
  \left\{
    \theta(\om-s\om_0)\delta[\om-s\om_0-(\xi+\eta)]
  \right.
  \nonumber\\
  \left.
    +
    \theta(-\om-s\om_0)\delta[\om+s\om_0-(\xi+\eta)]
  \right\}
  \,.
\end{eqnarray}
Here $\theta(x)$ is the Heaviside step function and we have used the definition
\begin{equation}\label{eq:f_sc}
  f(k,\xi,s) 
  =
  \frac{(2g^2)^s}{s!}
  +
  \frac{(g^2)^s}{s!}
  \left[
    2\left(\frac{\xi}{\tilde{W}}\right)^2 -1 
  \right]
  +
  \frac{(g^2)^s}{s!}\cos(2k)
  \,.
\end{equation}
The coherent part of the spectrum, non-zero for $|\om|<\om_0$, is given by
\begin{equation}\label{eq:sc_coherent}
  A^\mathrm{c}(k,\om)
  =
  z_k \delta[\om - (E_k + \eta)]
\end{equation}
with the renormalized band energy $E_k$ being the solution of
\begin{equation}\label{eq:sc_band}
  E_k 
  =
  \xi_k + \mathrm{Re}\,\Sigma(k,E_k+\eta)
  \,,
\end{equation}
where $\xi_k=-\tilde{W}\cos{k}$, and the spectral weight takes the form
\begin{eqnarray}
  \fl
  z_k^{-1}
  =
  \bigg|
  1 + 
  \frac{\tilde{W}}{2\pi}
  \sum_{s\geq1}\mathcal{P}
  \int_{-\tilde{W}}^{\tilde{W}}\frac{\rmd\xi}{\sqrt{1-(\xi/\tilde{W})^2}}
  f(k,\xi,s)
  \nonumber\\
  \times
  \left[
    \frac{\mathrm{\theta(\xi+\eta)}}{(E_k-s\om_0-\xi)^2}
    +
    \frac{\theta(-\xi-\eta)}{(E_k+s\om_0-\xi)^2}
  \right]
  \bigg|
  \,.
\end{eqnarray}

The imaginary part of the self-energy, determining the incoherent
excitations, is non-zero only for $|\om|>\om_0$. Consequently, we get
\begin{eqnarray}\label{eq:im_si_sc}
  \fl
  \mathrm{Im}\,\Sigma(k,\om\lessgtr\mp\om_0)
  =
  -\frac{\tilde{W}}{2}\sum_{s\geq1}
  \theta(\mp\om-s\om_0)
  \frac{f^{\mp}(k,\xi,s)}
  {(X^\pm_s)^\frac{1}{2}}
  \int_{-\tilde{W}}^{\tilde{W}}\rmd\xi \delta(\om\pm s\om_0-\eta-\xi)
  \,,
\end{eqnarray}
with
\begin{equation}
  X^\pm_s
  =
  1-\left(\frac{\om\pm s\om_0-\eta}{\tilde{W}}\right)^2
  \,.
\end{equation}
If $\om_0>2\tilde{W}$, for a given $\om$, only one term of the sum in
equation~(\ref{eq:im_si_sc}) contributes. The corresponding index $\sigma$ is
determined by the conditions
\begin{equation}
  \theta(\phantom{-}\om-\sigma\om_0) = 1
  \quad \mathrm{and} \quad
  \int_{-\tilde{W}}^{\tilde{W}}\rmd\xi \delta(\om-\sigma\om_0-\eta-\xi) = 1
\end{equation}
or
\begin{equation}
  \theta(-\om-\sigma\om_0) = 1
  \quad \mathrm{and} \quad
  \int_{-\tilde{W}}^{\tilde{W}}\rmd\xi \delta(\om+\sigma\om_0-\eta-\xi) = 1
  \,.
\end{equation}
The incoherent spectrum then consists of non-overlapping parts:
\begin{eqnarray}
  \fl
  A^\mathrm{ic}(k,\om\lessgtr\mp\om_0)
  =
  \frac{\tilde{W}}{2\pi}
  \frac{(X^\pm_\sigma)^{\frac{1}{2}} f^{\mp}(k,\om,\sigma)}
  {%
  \left[
    X^\pm_\sigma
    \left(\om-\xi_k-\eta-\mathrm{Re}\,\Sigma(k,\om)\right)^2
    +
    \left(
      \frac{\tilde{W}}{2} f^{\mp}(k,\om,\sigma)
    \right)^2
  \right]}
\,.
\end{eqnarray}
Here we have defined
\begin{equation}\label{eq:fpm}
  f^{\pm}(k,\om,s)
  =
  f(k,\xi=\om\mp s\om_0-\eta,s)
  \,.
\end{equation}

The polaron spectral function determines the equilibrium properties of the spinless
HM in the SC regime. The photoemission spectra, however, are
determined by the electron spectral function which is related to the retarded
Green function containing electronic operators.  According to the canonical
Lang-Firsov transformation~\cite{LangFirsov}, which defines small polaron
states, the relation between the polaronic operators $c_i$ entering the
Hamiltonian~(\ref{eq:HM}) with the coefficients~(\ref{eq:scC}), and the
transformed electron operators $\tilde{c}_i$, reads
\begin{equation}
  \tilde{c}_j^{\nag} = \exp[g(b^\dag_j-b^{\nag}_j)] c^{\nag}_j
  \,,\quad
  \tilde{c}^\dag_j = \exp[-g(b^\dag_j-b^{\nag}_j)] c^\dag_j
\end{equation}
or, in the Bloch representation,
\begin{equation}\fl
  \tilde{c}^{\nag}_k 
  =
  \frac{1}{\sqrt{N}}\sum_j \rme^{-\rmi k R_j} \exp[g(b^\dag_j - b^{\nag}_j)]
  c^{\nag}_j
  \,,\quad
  \tilde{c}^\dag_k 
  =
  \frac{1}{\sqrt{N}}\sum_j \rme^{\rmi k R_j} \exp[-g(b^\dag_j - b^{\nag}_j)]
  c^\dag_j
  \,.
\end{equation}

We start our derivation of the relation between the polaronic and electronic
spectra from the time-ordered Green function~\cite{Ma90} for the
(transformed) electron operators
\begin{equation}\fl
  \tilde{G}^\mathrm{T}(k,t_1,t_2) 
  =
  -\rmi\las T_t \tilde{c}^{\nag}_k(t_1)^{\nag} \tilde{c}_k^\dag(t_2)\ras
  =
  -\rmi\sum_d \rme^{\rmi k d} \las T_t \tilde{c}^{\nag}_j(t_1) \tilde{c}_{j+d}^\dag(t_2)\ras
\end{equation}
and factorize the statistical averages with respect to polaron and phonon variables
\begin{equation}\label{GTk}
  \fl
  \tilde{G}^\mathrm{T}(k,t_1,t_2)
  =
  \sum_d \rme^{\rmi k d} G^\mathrm{T}(t,d)
  \las 
  T_t \rme^{g[b^\dag_j(t)-b^{\nag}_j(t)]} \rme^{-g[b^\dag_{j+d}(0)-b^{\nag}_{j+d}(0)]}
  \ras
  \,,
\end{equation}
where $t=t_1-t_2$ and  $G^\mathrm{T}(d,t)=-\rmi\las T_t c^{\nag}_j(t)
c^\dag_{j+d}(0)\ras$ represents the time-ordered Green function of polaron
operators fulfilling
\begin{equation}
  \fl
  G^\mathrm{T}(d,t) 
  =
  N^{-1}\sum_{k'} \rme^{-\rmi k'd} G^\mathrm{T}(k',t)
  \,,\quad 
  G^\mathrm{T}(k',t) 
  =
  -\rmi\las T_t c^{\nag}_{k'}(t) c^\dag_{k'}(0)\ras
  \,.
\end{equation}
The averages over the phonon variables in equation~(\ref{GTk}) will be evaluated using 
mutually independent local Einstein oscillators having the time-dependence
$b_j(t)=\rme^{-\rmi\om_0t}b_j$. Working in the low-temperature approximation
we obtain
\begin{eqnarray}
  \fl
  \tilde{G}^\mathrm{T}(k,\om)
  =
  \rme^{-g^2} G^\mathrm{T}(k,\om)
  +
  \rme^{-g^2}\sum_{s\geq1}\frac{(g^2)^s}{s!}\frac{1}{N}
  \\\nonumber
  \times
  \sum_{k'}
  \left[
    \int_0^\infty\,\rmd t G^>(k',t) \rme^{\rmi(\om-s\om_0+\rmi\epsilon)t}
    +
    \int_{-\infty}^0\,\rmd t G^<(k',t) \rme^{\rmi(\om+s\om_0-\rmi\epsilon)t}
  \right]
\end{eqnarray}
with $G^>(k,t)=-\rmi\las c^{\nag}_k(t) c^{\dag}_{k}(0)\ras$ and
$G^<(k,t)=\rmi\las c^{\dag}_k(0) c^{\nag}_{k}(t)\ras$, and the convergence
factor $\exp(-\epsilon|t|)$, $\epsilon\rightarrow 0^+$.
Introducing the generalized function $\zeta(\omega)
= [\omega+\rmi\epsilon]^{-1}$, $\epsilon\rightarrow 0^+$, we get
\begin{eqnarray}
  \fl
  \tilde{G}^\mathrm{T}(k,\om)
  =
  \rme^{-g^2}G^\mathrm{T}(k,\om) + \rme^{-g^2}
  \sum_{s\geq1}\frac{(g^2)^s}{s!}\frac{1}{N}
  \sum_{k'}   
   \left[
    \int_{-\infty}^\infty\rmd\om' G^>(k',\om')
  \right.
  \nonumber\\
    \left.
      \times
      \rmi
    \zeta(\om-s\om_0-\om')
    -
    \int_{-\infty}^\infty\rmd \om' G^<(k',\om')\rmi
    \zeta^*(\om+s\om_0-\om')
  \right]\,.
\end{eqnarray}
The Green functions $G^{\gtrless}(k,\om)$ are related to the polaron
spectral function $\Ap(k,\om)$ through
\begin{equation}\fl
  G^<(k,\om) = \rmi n_\mathrm{F}(\om) \Ap(k,\om)
  \,,\quad
  G^>(k,\om) = -\rmi [1- n_\mathrm{F}(\om)] \Ap(k,\om)
  \,.
\end{equation}
At $T=0$, we have
\begin{eqnarray}
  \fl
  \tilde{G}^\mathrm{T}(k,\om)
  =
  \rme^{-g^2}G^\mathrm{T}(k,\om) + \rme^{-g^2}
  \sum_{s\geq1}\frac{(g^2)^s}{s!}\frac{1}{N}
  \sum_{k'}\left[
    \int_0^\infty\rmd \om'
    \Ap(k',\om') \zeta(\om-s\om_0-\om')
     \right.
    \nonumber\\
    +
    \left.
    \int_{-\infty}^0\rmd \om' 
    \Ap(k',\om') \zeta^*(\om+s\om_0-\om')
  \right]
  \,.
\end{eqnarray}
The relation between the time-ordered Green function $G^\mathrm{T}(k,\om)$
and the associated retarded Green function $G^\mathrm{R}(k,\om)$ in the
low-temperature approximation reads
\begin{equation}\label{rGRGK}
  \Im G^\mathrm{R}(k,\om) = \frac{\om}{|\om|}\Im G^\mathrm{T}(k,\om)
\end{equation}
and hence we obtain
\begin{equation}\label{rAGT}
  \Ap(k,\om) = -\frac{1}{\pi}\frac{\om}{|\om|}\Im G^\mathrm{T}(k,\om)
  \,.
\end{equation}
Of course, equations~(\ref{rGRGK}) and~(\ref{rAGT}) hold for the electron Green
function as well. Consequently, the electron spectral function $\Ae(k,\om)$
is expressed in terms of the polaron spectral function $\Ap(k,\om)$ as
\begin{eqnarray}\label{eq:sc_el_spectrum}
  \fl
  \Ae(k,\om)
  =
  \rme^{-g^2}\Ap(k,\om) 
  + \rme^{-g^2}\frac{1}{N}
  \sum_{s\geq1}\frac{(g^2)^s}{s!}
  \nonumber\\
  \hspace*{-1em}
  \times
  \sum_{k'}\left[
    \Ap(k',\om-s\om_0) 
    \theta(\om-s\om_0)
    +
    \Ap(k',\om+s\om_0)
    \theta(-\om-s\om_0)
  \right]
\,.
\end{eqnarray}
Similar results have been derived before in \cite{AlRa92,Al_book}.

\subsubsection{Intermediate coupling}\label{sec:interm-coupl}

As we shall see in section~\ref{sec:results}, the results of the WC
(SC) approximation are in good agreement with numerical calculations
\cite{HoNevdLWeLoFe04,HoWeAlFe05} if $\lambda,g\ll1$ ($\lambda,g\gg1$).
However, the cross-over between these limiting cases, revealed by the
numerical calculations, appears to be out of reach for the analytical
formulae hitherto deduced.

To interpolate between WC and SC, we shall modify the method of
canonical transformation by Lang and Firsov \cite{LangFirsov}. In the latter,
the term of the HM~(\ref{eq:HM}) linear in the local oscillator coordinate
$x_i=x_0(b^\dag_i + b^{\nag}_i)$ is completely eliminated by the
translational transformation $U=\exp[\sum_i g c^\dag_i c^{\nag}_i (b^\dag_i -
b^{\nag}_i)]$. As a result, the local lattice oscillator at the site $i$ is
shifted by $\Delta x_i=2x_0 g c^\dag_i c^{\nag}_i$, if occupied by a charge
carrier, whereas there is no such deformation at unoccupied sites.  To
generalize this picture, we abandon the site localization of both the charge
carrier and the lattice deformation in the transformation. Physically, these
localizations will be destroyed with increasing hopping rate and
charge-carrier concentration in the IC regime. 
Different canonical transformations, taking into account charge-density-wave
order at $n=0.5$, have been proposed, \eg, by Zheng \etal \cite{Zheng}. However,
in this approach, there is no filling dependence of their variational
parameter and of the mean lattice deformation, which is
crucial for a correct description of the adiabatic IC regime.

The probability for the charge carrier to be found at a distance $|r_j-r_i|$
from the center $r_i$ of the polaron will be assumed to be proportional to
$\exp(-|r_j-r_i|/R)$. In one dimension, and setting the lattice constant to
unity, we use the normalized distribution

\begin{equation}
  p(|j-i|) 
  =
  \rme^{-\frac{|j-i|}{R}}
  \tanh\frac{1}{2R}
  \,.
\end{equation}
Accordingly, the shift of the local oscillator with coordinate $x_i$ is
assumed to be
\begin{equation}\label{eq:shift}
  \Delta x_i 
  =
  2 x_0 g K c^\dag_i c^{\nag}_i + 2 x_0 g \gamma (1- c^\dag_i c^{\nag}_i)
\end{equation}
with
\begin{equation}
  K 
  =
  \tanh\frac{1}{2R}
  \,,
  \quad
  \gamma 
  = 
  2 K \rme^{-\frac{1}{R}} n
  \,.
\end{equation}
The last term in equation~(\ref{eq:shift}), characterizing the mean lattice
deformation background, takes into account the influence of nearest-neighbor
sites only.

The canonical transformation leading to the oscillator shift~(\ref{eq:shift})
reads
\begin{equation}\label{eq:ic_transformation}
  U 
  =
  \exp\left[\sum_i g (\gammab c^\dag_i c^{\nag}_i + \gamma) (b^\dag_i - b^{\nag}_i)\right]
  \,,\quad
  \gammab 
  =
  K - \gamma
  \,.
\end{equation}
Carrying out the transformation $U^\dag H U=\tilde{H}$ for the
Hamiltonian~(\ref{eq:HM}) with~(\ref{eq:wcC}), the terms of $\tilde{H}$ containing polaron
operators are modified with the following coefficients [cf.
equations~(\ref{eq:wcC}) and (\ref{eq:scC})]
\begin{equation}\label{eq:trans_ic}
  C_{ii}
  =
  g\omega_0 (1-\gammab)(b^\dag_i+b^{\nag}_i)
  \,,\quad
  C_{\las ij \ras}
  =
  t \rme^{-\gammab g (b^\dag_i - b^{\nag}_i - b^\dag_j + b^{\nag}_j)}
  \,.
\end{equation}
Moreover, we have $\eta=-\mu -\Ep [\gammab (2-\gammab) + 2\gamma
(1-\gammab)]$.

Owing to numerical problems which occur in certain parameter regimes when
using the full Hamiltonian with the coefficients~(\ref{eq:trans_ic}), the
variational parameter $R$ of the transformation will be determined in the
first approximation, which is analogous to the Hartree approximation. The
corresponding polaron spectral function is given as
$\Ap(k,\om)=\delta[\om-(\xi_k+\eta)]$, where $\xi_k=-\tilde{W}\cos k$ with
$\tilde{W}=W\exp[-(\gammab g)^2]$, and $\eta$ as defined above. $R$ is then
defined by the position of the minimum of the total energy per site $E/N$ in
the first approximation, \ie,
\begin{equation}\label{eq:energ}
  \frac{E}{N}
  =
  \frac{1}{\pi\tilde{W}}
  \int_{-\tilde{W}}^{\tilde{W}}\,\rmd \xi
  \frac{\xi+\eta}{\sqrt{1-(\xi/\tilde{W})^2}}
  \theta(-\xi-\eta) + \mu n + \Ep \gamma^2
  \,,
\end{equation}
with the condition for $\mu$
\begin{equation}
  \frac{1}{\pi\tilde{W}}
  \int_{-\tilde{W}}^{\tilde{W}}\,\rmd \xi
  \frac{\theta(-\xi-\eta)}{\sqrt{1-(\xi/\tilde{W})^2}}
  =
  n
  \,.
\end{equation}
The last term in equation~(\ref{eq:energ}) arises in $\tilde{H}$ from the lattice
deformation background at finite concentration $n$.

$R$ determined in this way for each set of model parameters will be used to
calculate both the electron and polaron spectral function, taking into
account the multi-phonon processes included in $\tilde{H}$ [cf.
equation~(\ref{eq:trans_ic})].

Using the same procedure as in the SC case, we calculate the self-energy and
spectral function at $T=0$, finding
\begin{eqnarray}
  \fl
  \mathrm{Re}\,\Sigma(k,\om)
  =
  \frac{\tilde{W}}{2\pi}\sum_{s\geq1}
  \mathcal{P}\int_{-\tilde{W}}^{\tilde{W}}\frac{\rmd\xi}{\sqrt{1-(\xi/\tilde{W})^2}}
  \overline{f}(k,\xi,s)
  \nonumber\\
  \times
  \left[
    \frac{\theta(\xi+\eta)}{\om-s\om_0-(\xi+\eta)}
    +
    \frac{\theta(-\xi-\eta)}{\om+s\om_0-(\xi+\eta)}
  \right]
  \nonumber\\
  \lo- 
  2\gammab(1-\gammab)\frac{\Ep}{\pi}
  \mathcal{P}\int_{-\tilde{W}}^{\tilde{W}}\frac{\rmd\xi}{\sqrt{1-(\xi/\tilde{W})^2}}
  \left(\frac{\xi}{\tilde{W}}+\cos{k}\right)
  \nonumber\\
  \times
  \left[
    \frac{\theta(\xi+\eta)}{\om-\om_0-(\xi+\eta)}
    -
    \frac{\theta(-\xi-\eta)}{\om+\om_0-(\xi+\eta)}
  \right]
  \nonumber\\
  \lo+
  (1-\gammab)^2\frac{\Ep\omega_0}{\pi\tilde{W}}
  \mathcal{P}\int_{-\tilde{W}}^{\tilde{W}}\frac{\rmd\xi}{\sqrt{1-(\xi/\tilde{W})^2}}
  \nonumber\\
  \times
  \left[
    \frac{\theta(\xi+\eta)}{\om-\om_0-(\xi+\eta)}
    +
    \frac{\theta(-\xi-\eta)}{\om+\om_0-(\xi+\eta)}
  \right]
\end{eqnarray}
and
\begin{eqnarray}\label{eq:im_si_ic}
  \fl
  \mathrm{Im}\,\Sigma(k,\om\lessgtr\mp\om_0)
  =
  -\frac{\tilde{W}}{2}\sum_{s\geq1}
  \theta(\mp\om-s\om_0)
  \frac{\overline{f}^{\mp}(k,\om,s)}
  {(X^\pm_s)^\frac{1}{2}}
  \int_{-\tilde{W}}^{\tilde{W}}\rmd\xi \delta(\om\pm s\om_0-\eta-\xi)
  \nonumber\\
  \lo\mp
  2\gammab(1-\gammab)\Ep (X^\pm_1)^{-\frac{1}{2}}
  \left(\frac{\omega\pm\omega_0-\eta}{\tilde{W}}+\cos{k} \right)
  \nonumber\\
  \times
  \theta(\mp\om-\om_0)
  \int_{-\tilde{W}}^{\tilde{W}}\rmd\xi \delta(\om\pm\om_0-\eta-\xi)
  \nonumber\\
  \lo-
  (1-\gammab)^2 \frac{\Ep\omega_0}{\tilde{W}} (X^\pm_1)^{-\frac{1}{2}}
  \theta(\mp\om-\om_0)
  \int_{-\tilde{W}}^{\tilde{W}}\rmd\xi \delta(\om\pm\om_0-\eta-\xi)
  \,.
\end{eqnarray}
Here $\tilde{W}=W\rme^{-(\gammab g)^2}$ and
$\overline{f}$, $\overline{f}^\pm$ are defined as in equations~(\ref{eq:f_sc}) and
(\ref{eq:fpm}) but with $g$ replaced by $g\gammab$. Note that the above
equations reproduce the corresponding SC results in the limit $R=0$
($\gammab=1$), and also the WC ones in the limit $R=\infty$ ($\gammab=0$).

The coherent spectrum is again given by equations~(\ref{eq:sc_coherent}) and
(\ref{eq:sc_band}), with the spectral weight
\begin{equation}
  z_k^{-1}
  =
  \left|
   1  - [\partial\mathrm{Re} \Sigma(k,\omega) / \partial\omega]_{\omega=E_k+\eta}
  \right|
  \,,
\end{equation}
whereas the incoherent part takes the familiar form
\begin{equation}
  \Ae^\mathrm{ic}(k,\omega)
  =
  -\frac{1}{\pi}
  \frac{\mathrm{Im}\Sigma(k,\omega)}
  {[\omega-(\xi_k+\eta) - \mathrm{Re}\Sigma(k,\omega)]^2 +
    [\mathrm{Im}\Sigma(k,\omega)]^2}
  \,.
\end{equation}

Finally, equation~(\ref{eq:sc_el_spectrum}), which determines the relation between
the electronic and polaronic spectral functions, also applies to the present
case if $g$ is replaced by $g\gammab$ throughout.

\section{Numerical results}\label{sec:results}

%
\begin{figure}
  \begin{center}
  \includegraphics[width=0.495\textwidth]{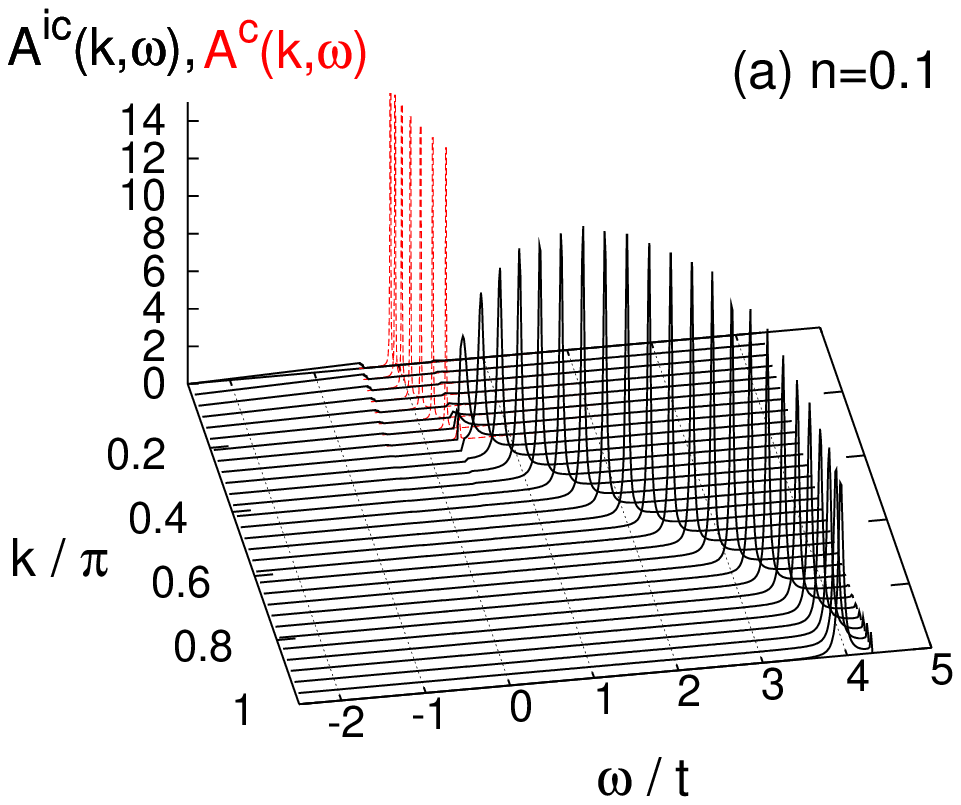}
  \includegraphics[width=0.495\textwidth]{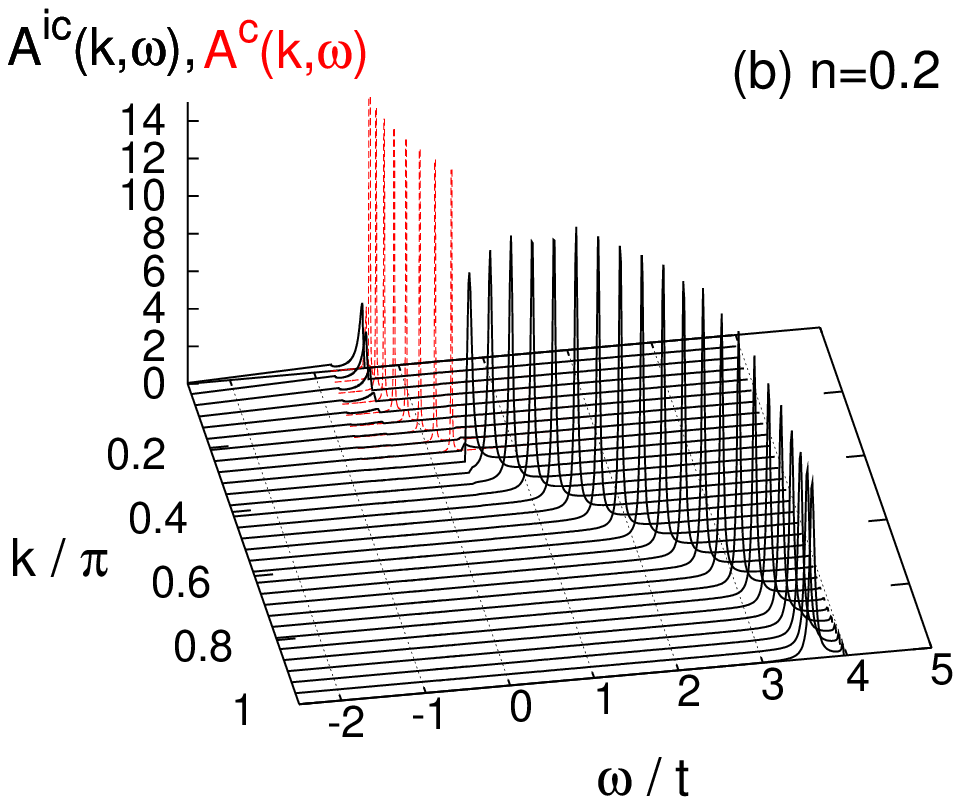}\\
  \includegraphics[width=0.495\textwidth]{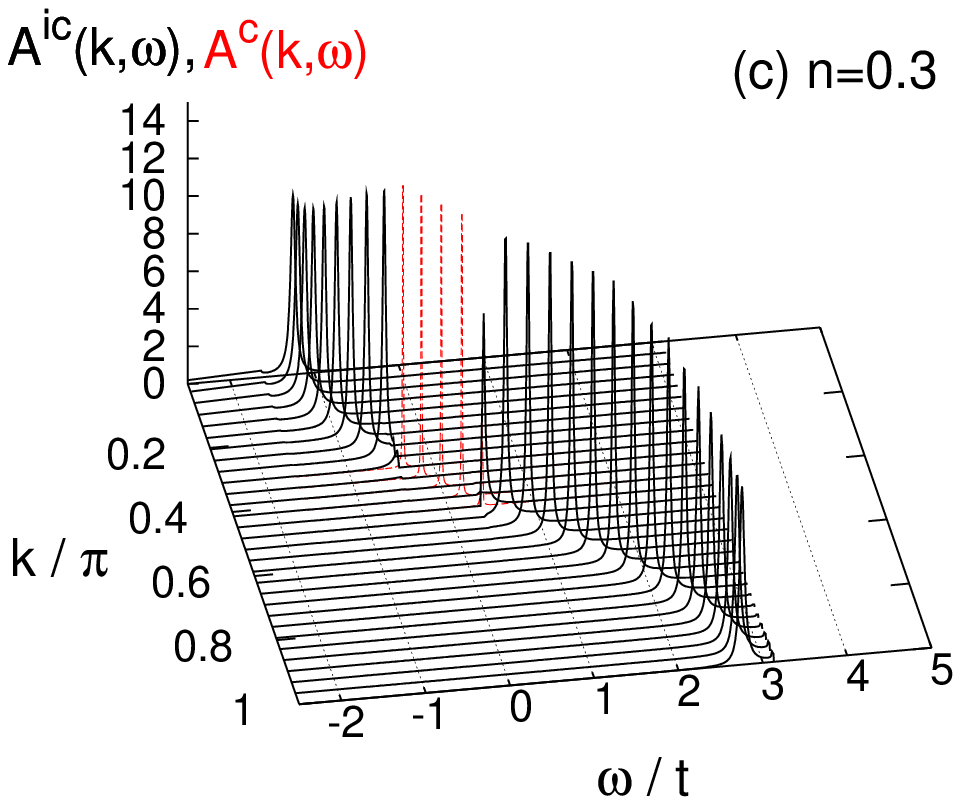}
  \includegraphics[width=0.495\textwidth]{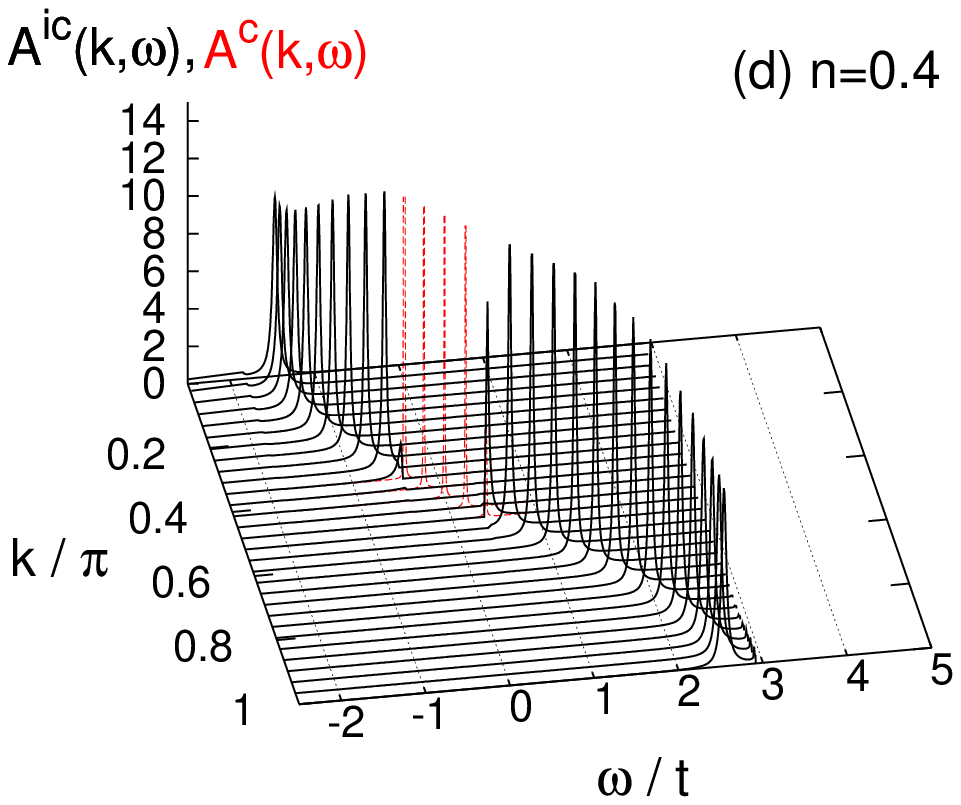}
  \end{center}
  \caption{\label{fig:wc_density_w0.4}%
    (colour online) Coherent ($A^\mathrm{c}$, $\dashed$) and incoherent
    ($A^\mathrm{ic}$, $\full$) parts of the spectral function in the
    weak-coupling approximation for different band fillings $n$. Here
    $\om_0/t=0.4$ and $\Ep/t=0.1$.}
\end{figure}

As in section~\ref{sec:GFapproach}, we first discuss the WC and SC limits,
before turning to the important IC regime. Since we use a finite number of
momenta $k$, it is not possible to tune the band filling $n$ (via the chemical
potential $\mu$) to a specific, desired value with arbitrary accuracy. In
order to simplify the discussion of the different density regimes, we
therefore report rounded values of $n$ in the figures and the text. The
largest deviations of the actual $n$ from the value reported occur in
the SC case for which, however, the density dependence is very weak (see
section~\ref{sec:res:strong-coupling}).

\subsection{Weak coupling}\label{sec:res:weak-coupling}

\begin{figure}
  \begin{center}
 \includegraphics[width=0.495\textwidth]{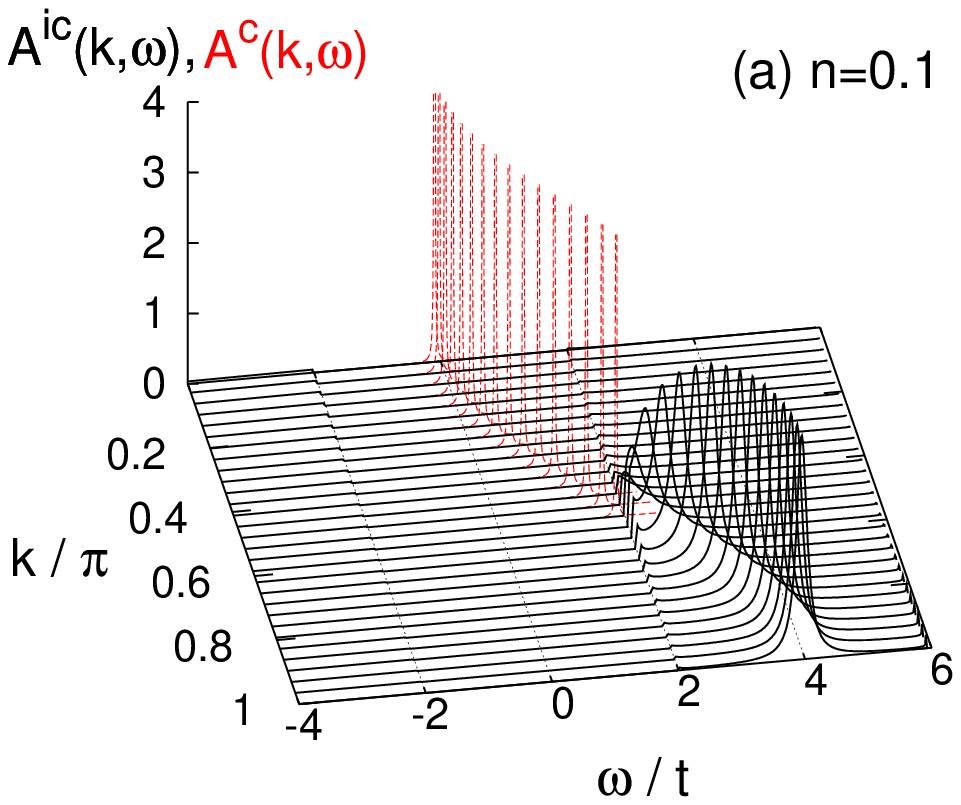}
 \includegraphics[width=0.495\textwidth]{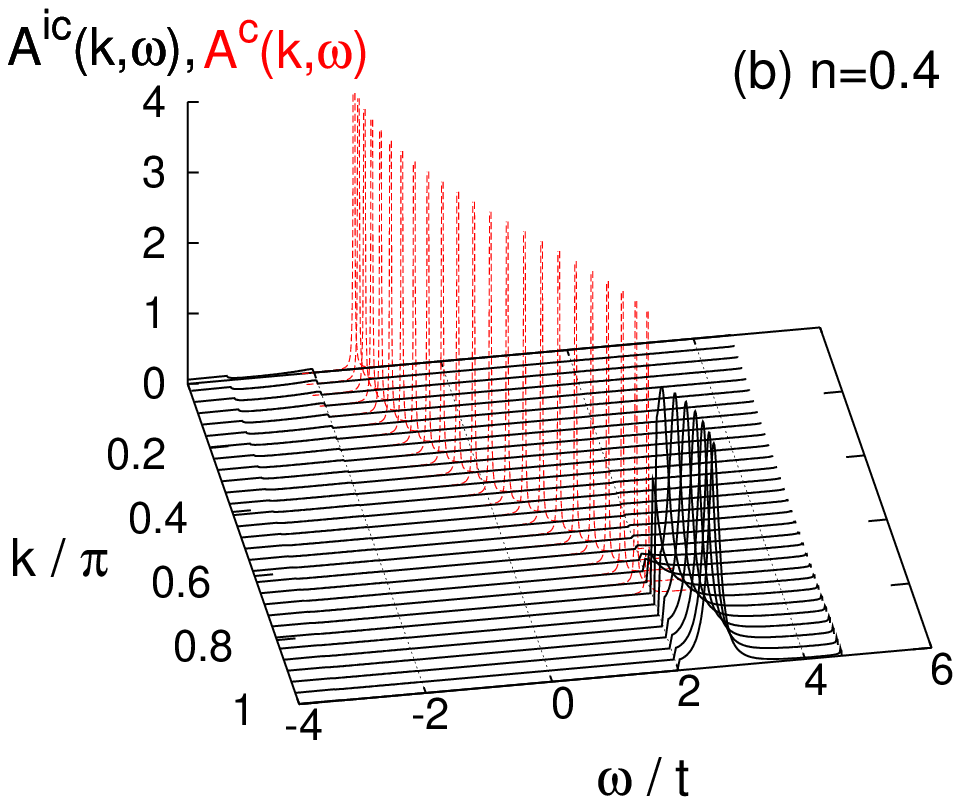}
  \end{center}
  \caption{\label{fig:wc_density_w2.0}%
    (colour online)
    As in figure~\ref{fig:wc_density_w0.4}, but for $\om_0/t=2$.
  }
\end{figure}

\begin{figure}
  \begin{center}
  \includegraphics[width=0.495\textwidth]{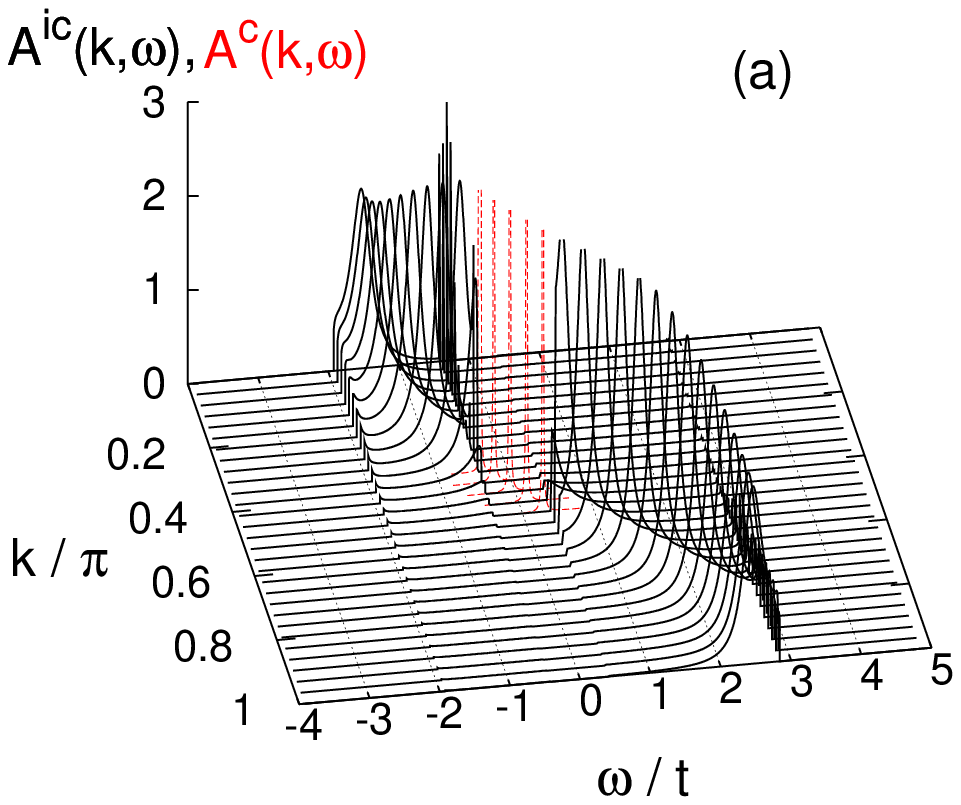}
  \includegraphics[width=0.495\textwidth]{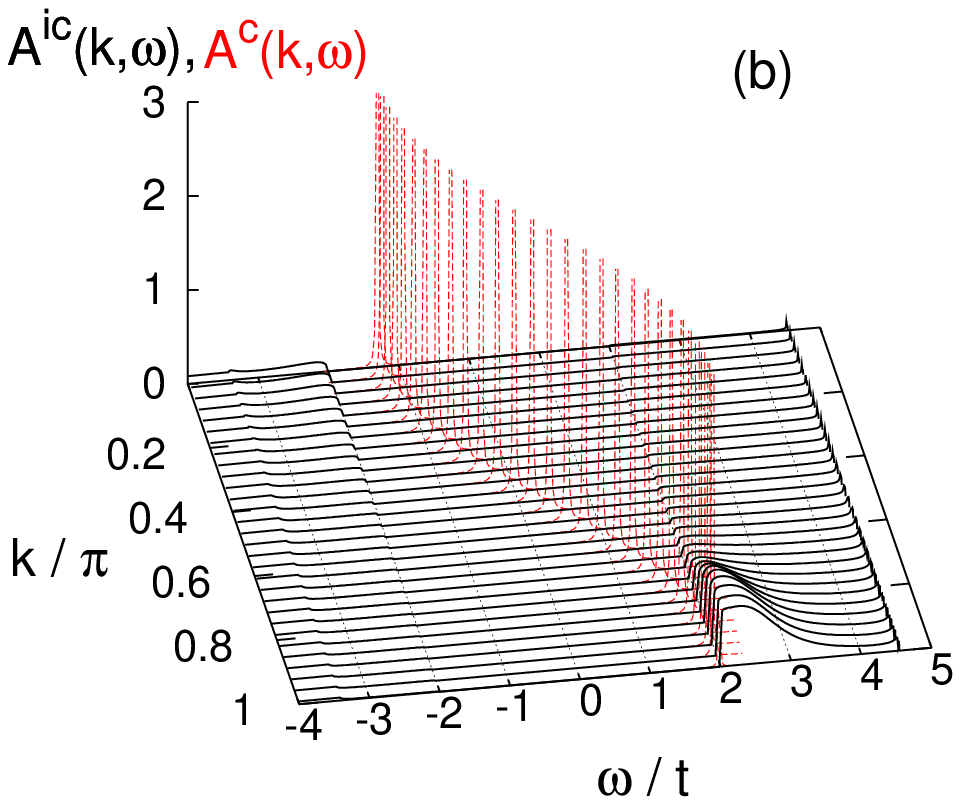}
  \end{center}
  \caption{\label{fig:coupling_wc}%
    (colour online) As in figure~\ref{fig:wc_density_w0.4}, but for $\Ep/t=0.5$
    and fixed band filling $n=0.4$. Here (a) $\om_0/t=0.4$ and (b)
    $\om_0/t=2$.  }
\end{figure}

Figures~\ref{fig:wc_density_w0.4}--\ref{fig:coupling_wc} show the electronic
spectral function $A(k,\om)$ obtained from the WC approximation. The coherent
spectrum ($\Im\Sigma\equiv0$) is given as the solution of
equations~(\ref{eq:wc_A_coh})--(\ref{eq:wc_zk}) with energies
$|E_k-\mu|<\omega_0$. For $|\om|>\omega_0$, $A^\mathrm{ic}$, calculated
according to equations~(\ref{eq:wc_A_inc1}) and~(\ref{eq:wc_A_inc2}), consists of
peaks having widths proportional to $\Ep$. A comparison of
figures~\ref{fig:wc_density_w0.4} (a) and~\ref{fig:wc_density_w2.0}(a) shows the
spreading of the coherent spectrum with increasing $\om_0$. Finally,
comparing figure~\ref{fig:wc_density_w0.4}(d) [\ref{fig:wc_density_w2.0}(b)]
with figure~\ref{fig:coupling_wc}(a) [\ref{fig:coupling_wc}(b)], we
observe a broadening of the peaks in $A^\mathrm{ic}$ with increasing $\Ep$.

The evolution of the spectral function with increasing carrier density $n$ is
illustrated in figure~\ref{fig:wc_density_w0.4}(a)--(d). The coherent part is shifted
inside the spectrum as a function of $n$ (\ie, with $\mu$). Additionally, the
shape of $A^\mathrm{ic}$ is also affected by $n$, due to the dependence of
equations~(\ref{eq:wc_A_inc1}) and~(\ref{eq:wc_A_inc2}) on the chemical potential
$\mu$.

In figure~\ref{fig:weight_wc}, we plot the total spectral weight $\int\rmd k
\int\rmd \om A(k,\om)$, as well as the coherent weight $\int\rmd k\int \rmd
\om A^\mathrm{c}(k,\om)$. Note that for any $\Ep>0$, the coherent band $E_k$
is restricted to the interval $[-\om_0,\om_0]$, so that the corresponding
coherent weight is significantly smaller than the $\Ep=0$ value of unity.
This reduction is less pronounced for larger phonon frequency
$\om_0/t=2\approx W/t$ (figure~\ref{fig:weight_wc}). Furthermore, we see that
with increasing $\Ep$, the sum rule for $A(k,\om)$ becomes more and more
violated, as expected for a WC approximation (for a more detailed discussion
of sum rules see section~\ref{sec:res:strong-coupling}).

\begin{figure}
  \begin{center}
  \includegraphics[width=0.495\textwidth]{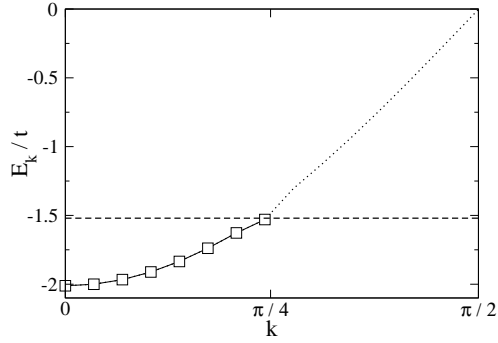}
  \end{center}
  \caption{\label{fig:Ek_wc}%
    Renormalized band $E_k$ for $\om_0/t=0.4$, $\Ep/t=0.1$ and $n=0.1$.
    Symbols indicate non-zero coherent spectral weight $z_k$. The horizontal
    line ($\dashed$) corresponds to $(E_0+\om_0)/t$.}
\end{figure}

Finally, figure~\ref{fig:Ek_wc} displays the coherent band dispersion $E_k$ at
small carrier density $n=0.1$. As known from the single-polaron problem, the
coherent weight $z_k$ drops to zero as the bare phonon
dispersion intersects with the renormalized band. This gives rise to a
flattening of the coherent band at large $k$~\cite{St96}, which is well
reproduced by the simple WC approximation.

\begin{figure}
  \begin{center}
 \includegraphics[width=0.495\textwidth]{fig4.eps}
  \end{center}
  \caption{\label{fig:weight_wc}%
    Coherent ($\dashed$) and total ($\full$) spectral weight from $A(k,\om)$ (see text) for
    $\om_0/t=0.4$ ($\opensquare$, $\fullsquare$) and $\om_0/t=2$
    ($\opencircle$, $\fullcircle$) as a
    function of the polaron binding energy $\Ep$. Here $\mu=0$.  }
\end{figure}

\subsection{Strong coupling}\label{sec:res:strong-coupling}

%
\begin{figure}
  \includegraphics[width=0.495\textwidth]{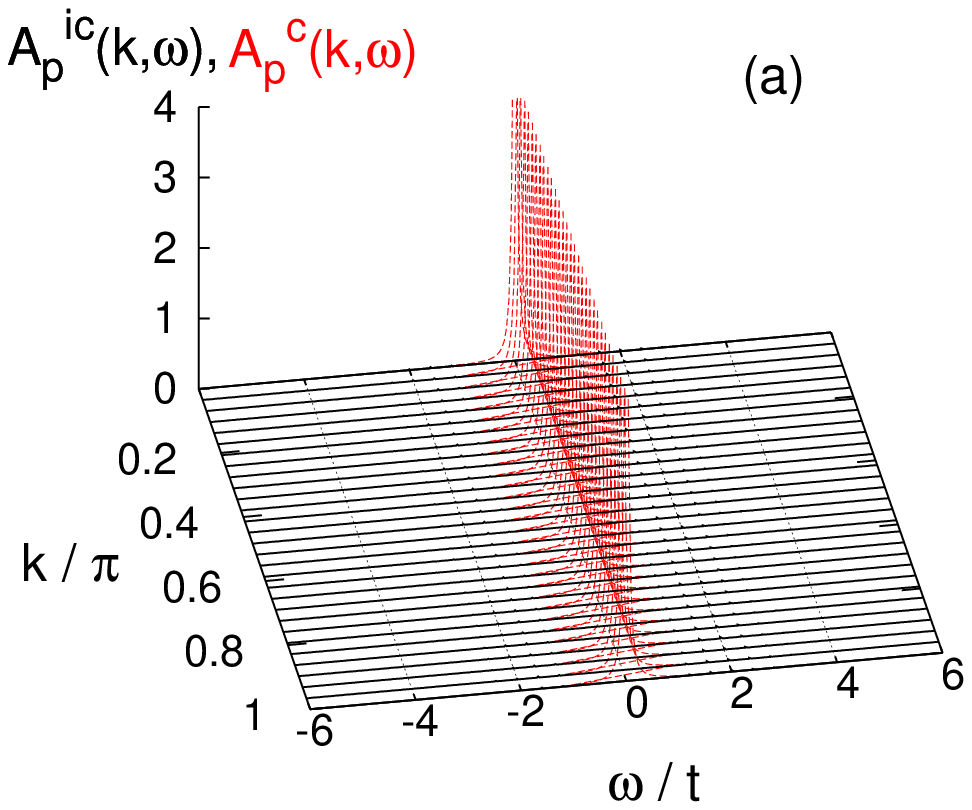}
  \includegraphics[width=0.495\textwidth]{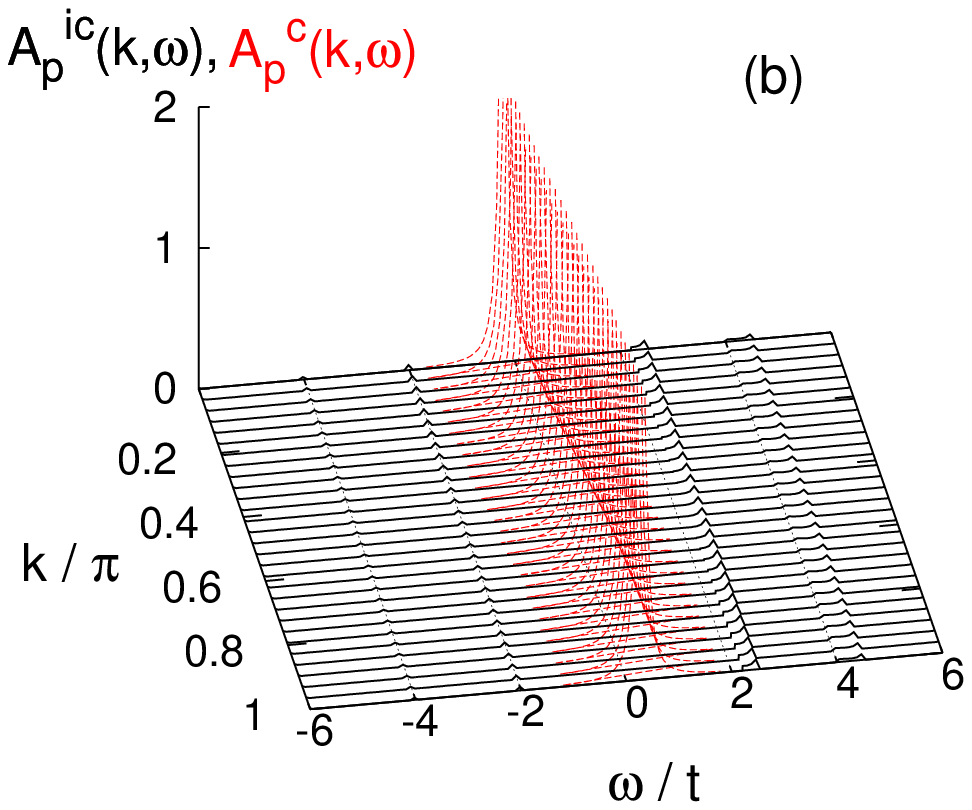}
  \caption{\label{fig:polaron_spectrum_sc}%
    (colour online)
    Coherent ($\Ap^\mathrm{c}$, $\dashed$) and incoherent ($\Ap^\mathrm{ic}$,
    $\full$) parts of the polaron spectral function in the strong-coupling
    approximation. Here $n=0.4$, $\Ep/t=4$ and (a) $\om_0/t=0.4$, (b)
    $\om_0/t=2$.
  }
\end{figure}
\begin{figure}
  \begin{center}
  \includegraphics[width=0.495\textwidth]{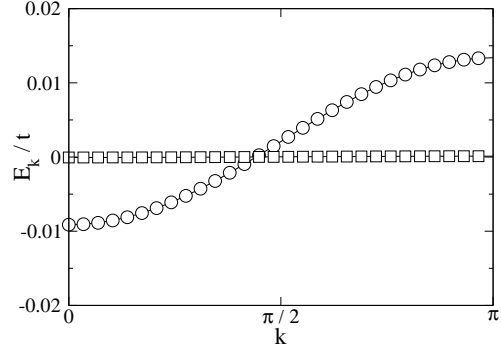}
  \end{center}
  \caption{\label{fig:Ek_sc}%
    Renormalized band $E_k$ for $n=0.4$, $\om_0/t=0.4$ and different
    values $\Ep/t=2$ ($\opencircle$) and $\Ep/t=4$ ($\opensquare$).}
\end{figure}
\begin{figure}
  \begin{center}
  \includegraphics[width=0.495\textwidth]{fig8.eps}
  \end{center}
  \caption{\label{fig:weight_sc}%
    Coherent (polaronic, $\dashed$) and total ($\full$) spectral weight
    from $\Ap(k,\om)$ for
    $\om_0/t=0.4$ ($\opensquare$, $\fullsquare$) and $\om_0/t=2$
    ($\opencircle$, $\fullcircle$), as a
    function of the polaron binding energy $\Ep$. Here $\mu=0$. 
}
\end{figure}
\begin{figure}
  \begin{center}
  \includegraphics[width=0.495\textwidth]{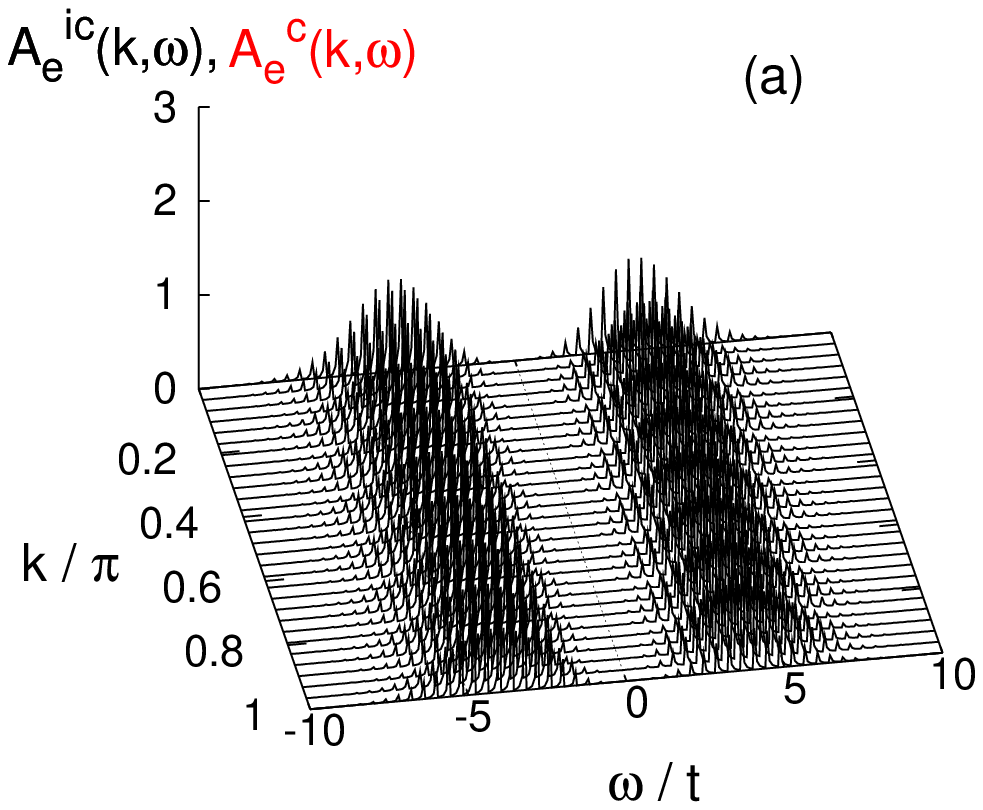}
  \includegraphics[width=0.495\textwidth]{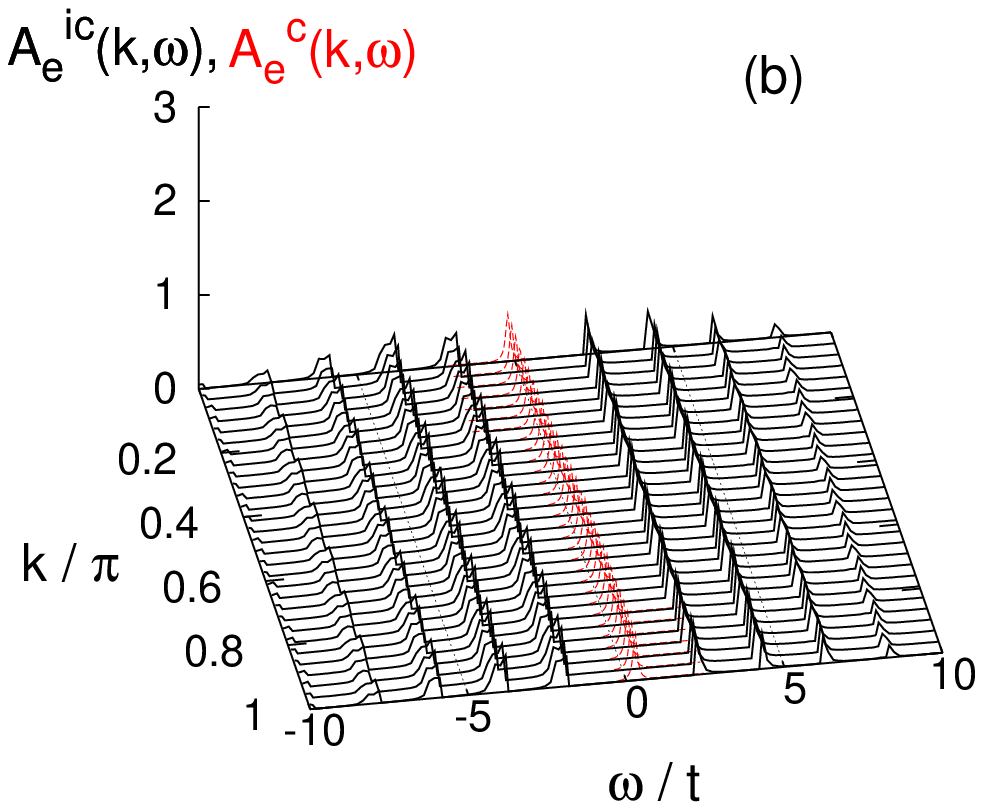}
  \end{center}
  \caption{\label{fig:coupling_sc}%
    (colour online)
    As in figure~\ref{fig:polaron_spectrum_sc}, but showing the electronic spectral
    functions $\Ae^\mathrm{c}$ ($\dashed$) and $\Ae^\mathrm{ic}$ ($\full$).}
\end{figure}
\begin{figure}
  \begin{center}
  \includegraphics[width=0.495\textwidth]{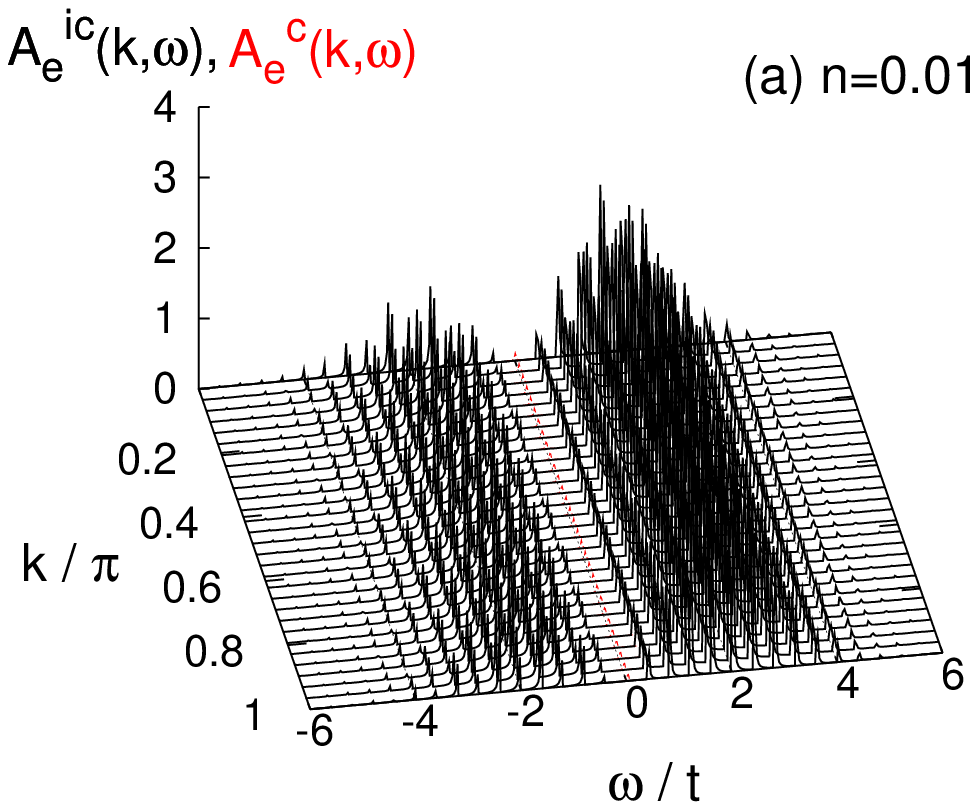}
  \includegraphics[width=0.495\textwidth]{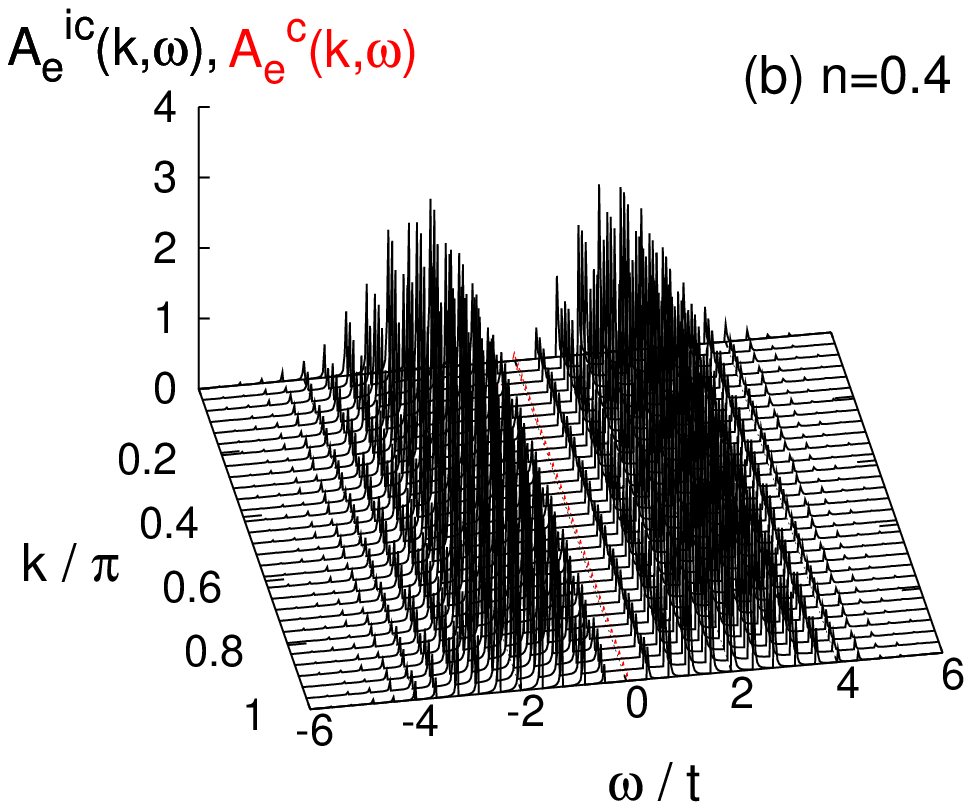}
  \end{center}
  \caption{\label{fig:sc_density_w0.4}%
    (colour online)
    As in figure~\ref{fig:coupling_sc}, but for different band fillings
    $n$. Here $\om_0/t=0.4$ and $\Ep/t=2$.}
\end{figure}
\begin{figure}
  \begin{center}
  \includegraphics[width=0.495\textwidth]{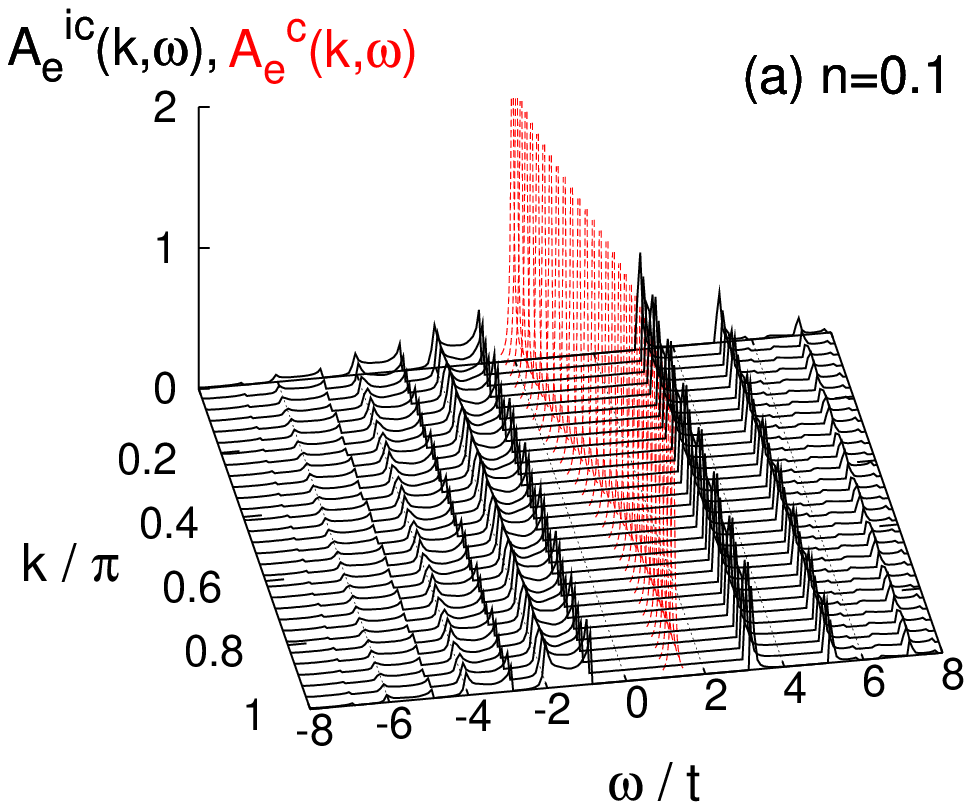}
  \includegraphics[width=0.495\textwidth]{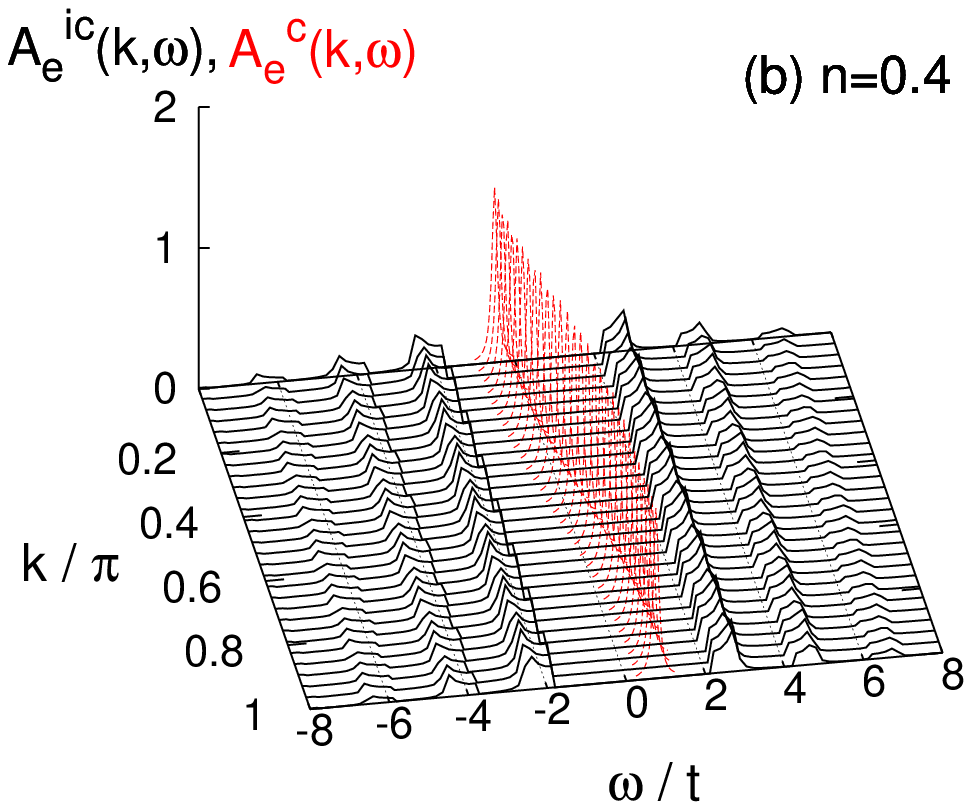}
  \end{center}
  \caption{\label{fig:sc_density_w2.0}%
    (colour online)
    As in figure~\ref{fig:sc_density_w0.4}, but for $\om_0/t=2$.}
\end{figure}

We now turn to the opposite, SC limit. The theory presented in
section~\ref{sec:strong-coupling} directly yields the polaronic spectrum
$\Ap(k,\om)$, results for which are shown in
figure~\ref{fig:polaron_spectrum_sc}(a) for $\om_0/t=0.4$ and $\Ep/t=4$. The
spectrum is dominated by a coherent polaronic band with negligible width (for
the dependence of the bandwidth on $\Ep$ see figure~\ref{fig:Ek_sc}) having a
spectral weight $z_k$ close to unity (cf. figure~\ref{fig:weight_sc}). This
suggests that small polarons are the correct quasiparticles in the SC regime.
Note that opposite to the WC case, where the sum rule for the spectral
function becomes more and more violated with increasing coupling
(figure~\ref{fig:weight_wc}), here the SC approximation becomes increasingly
better with increasing $\Ep$ (figure~\ref{fig:weight_sc}). 

We would like to point out that such changes in the total spectral weight are
absent in the work of Alexandrov and Ranninger~\cite{AlRa92}, since the latter was
restricted to the lowest (first) order of the self-energy, similar to the
Hartree approximation discussed in section~\ref{sec:interm-coupl}. In general,
the total spectral weight contained in the (electronic or polaronic) spectral
function for given parameters depends on the approximations made. As
illustrated by figures~\ref{fig:weight_wc} and~\ref{fig:weight_sc}, the total
spectral weight approaches the exact value of unity in the WC and SC regimes,
respectively, so that the normalization of the spectrum serves as a measure
of the validity of the underlying approximations. Since in
the present case even the first moment (\ie, the normalization) shows
deviations from exact results, we have refrained from checking the more
complicated sum rules derived in \cite{Korni}. This is also true of the IC
case discussed below.

The effect of increasing the phonon frequency $\om_0$ can be seen by
comparing figures~\ref{fig:polaron_spectrum_sc}(a) and (b). Most noticeably,
for larger $\om_0$, the width of the coherent polaron band---roughly scaling
proportional to $\rme^{-g^2}$---is larger.

The polaronic spectrum is related to the electronic spectrum $\Ae$ by
equation~(\ref{eq:sc_el_spectrum}), and typical results in the adiabatic and
non-adiabatic regimes are shown in figure~\ref{fig:coupling_sc}. Although
strictly speaking a distinction between coherent and incoherent contributions
cannot be made in the case of $\Ae$ [cf. equation~(\ref{eq:sc_el_spectrum})],
it is useful to separate the two terms on the r.~h.~s. of
equation~(\ref{eq:sc_el_spectrum}), and to identify the first as the
contribution of the coherent polaron band.

Carrying out the transformation from $\Ap$ to $\Ae$ according to
equation~(\ref{eq:sc_el_spectrum}), the weight of the coherent polaron band
visible in figure~\ref{fig:polaron_spectrum_sc} approximately acquires a
prefactor $\rme^{-g^2}$. The remaining contributions to the electronic
spectrum correspond to phonon-assisted photoemission processes.

In the case of $\Ae$, the main difference between the adiabatic
[figure~\ref{fig:coupling_sc}(a)] and the non-adiabatic regime
[figure~\ref{fig:coupling_sc}(b)] is the significantly larger weight of the
coherent band for large $\om_0$ since $g^2=\Ep/\om_0$. Consequently, the
weight contained in the incoherent excitations is noticeably reduced.

The results in figures~\ref{fig:sc_density_w0.4} and~\ref{fig:sc_density_w2.0}
show a certain dependence on the band filling $n$, but there occur no
qualitative changes even at IC $\Ep/t=2$. This is in contrast to recent
numerical work \cite{HoNevdLWeLoFe04,HoWeAlFe05}. In particular, the spectrum
in figure~\ref{fig:sc_density_w0.4}(b) is substantially different from
figures~\ref{fig:ic_w0.4}(c) and (d), which are all for the same
parameters. We shall see below that a more satisfactory description of the
real physics can be obtained using the variational approach discussed in
section~\ref{sec:interm-coupl}.

\subsection{Cross-over from weak to strong coupling}\label{sec:res:ic}

\begin{figure}
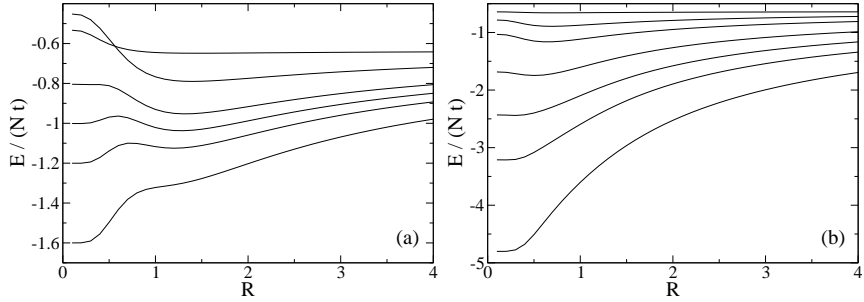

  \begin{center}
  \includegraphics[height=0.3\textwidth]{fig12a.eps}
  \includegraphics[height=0.3\textwidth]{fig12b.eps}
  \end{center}
  \caption{\label{fig:free_energy_ic}%
    Total energy per site $E/N$ as a function of the variational parameter $R$
    for different values of $\Ep$ from the Hartree approximation. Here
    $n=0.4$ and (a) $\om_0/t=0.4$, (b) $\om_0/t=2$. The values of $\Ep/t$ are
    (from top) (a) 0.1, 1, 2, 2.5, 3, 4, (b) 0.1, 1, 2, 4, 6, 8, and 12.
  }
\end{figure}
\begin{figure}
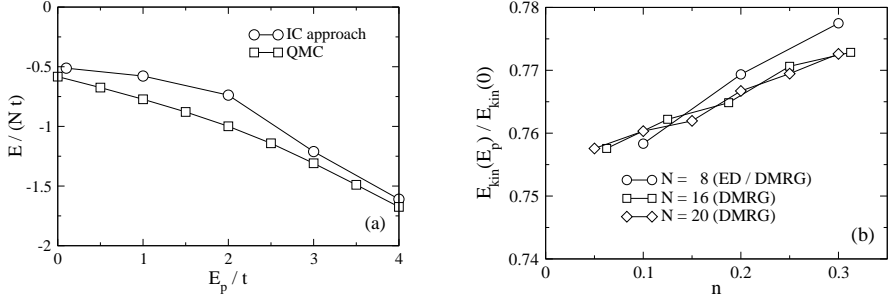

  \begin{center}
  \includegraphics[height=0.3\textwidth]{fig13a.eps}
  \hspace*{2em}
  \includegraphics[height=0.3\textwidth]{fig13b.eps}
  \end{center}
  \caption{\label{fig:E0_Ep}%
    (a) Total energy per site $E/N$ as a function of $\Ep$ obtained from
    $A(k,\om)$ (see text), where $n=0.4$ and $\om_0/t=0.4$. Also shown are
    quantum Monte Carlo results at inverse temperature $\beta t=8$ obtained with the method of
    \cite{HoNevdLWeLoFe04}. (b) Exact diagonalization (ED) and density matrix
    renormalization group (DMRG) results for the renormalized
    kinetic energy as a function of carrier density $n$ for $\Ep/t=2$,
    $\om_0/t=0.4$ and different cluster sizes $N$. 
  }
\end{figure}
\begin{figure}
  \includegraphics[width=0.495\textwidth]{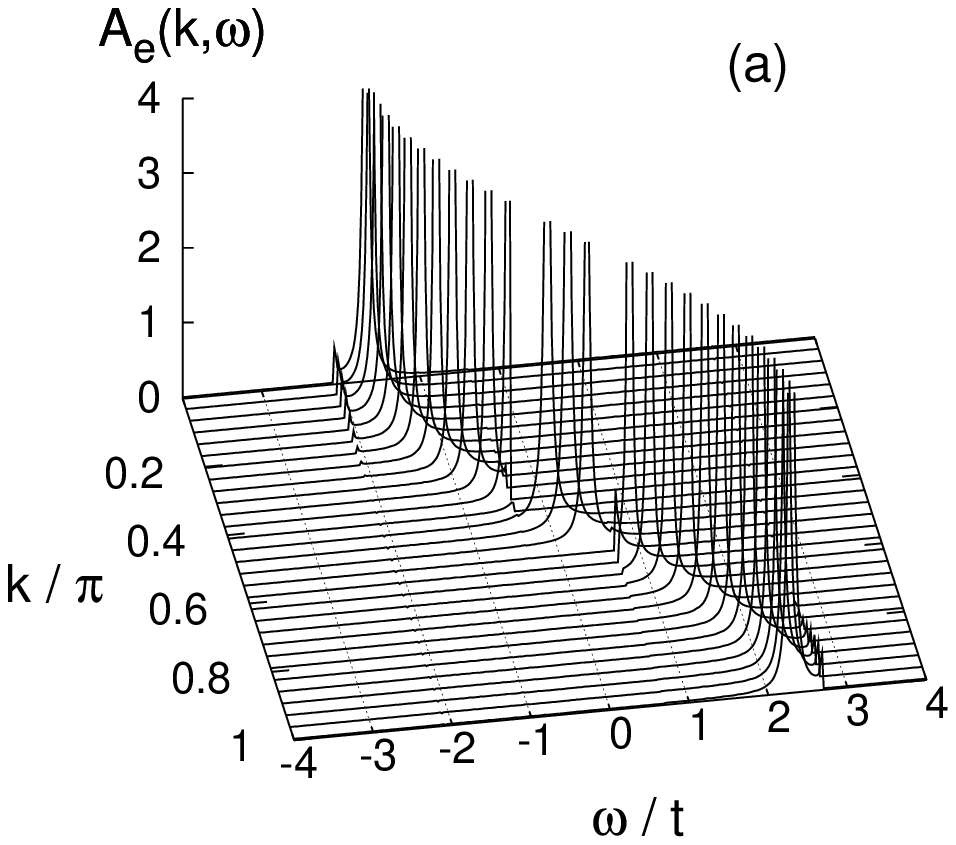}
  \includegraphics[width=0.495\textwidth]{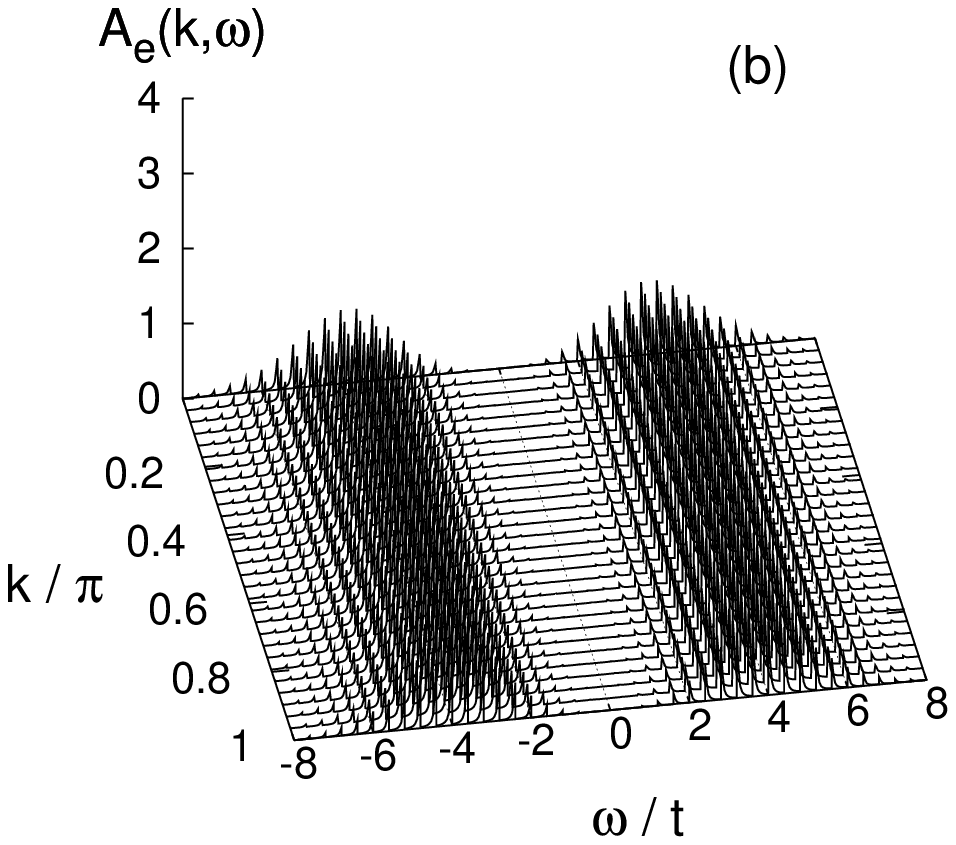}\\
  \includegraphics[width=0.495\textwidth]{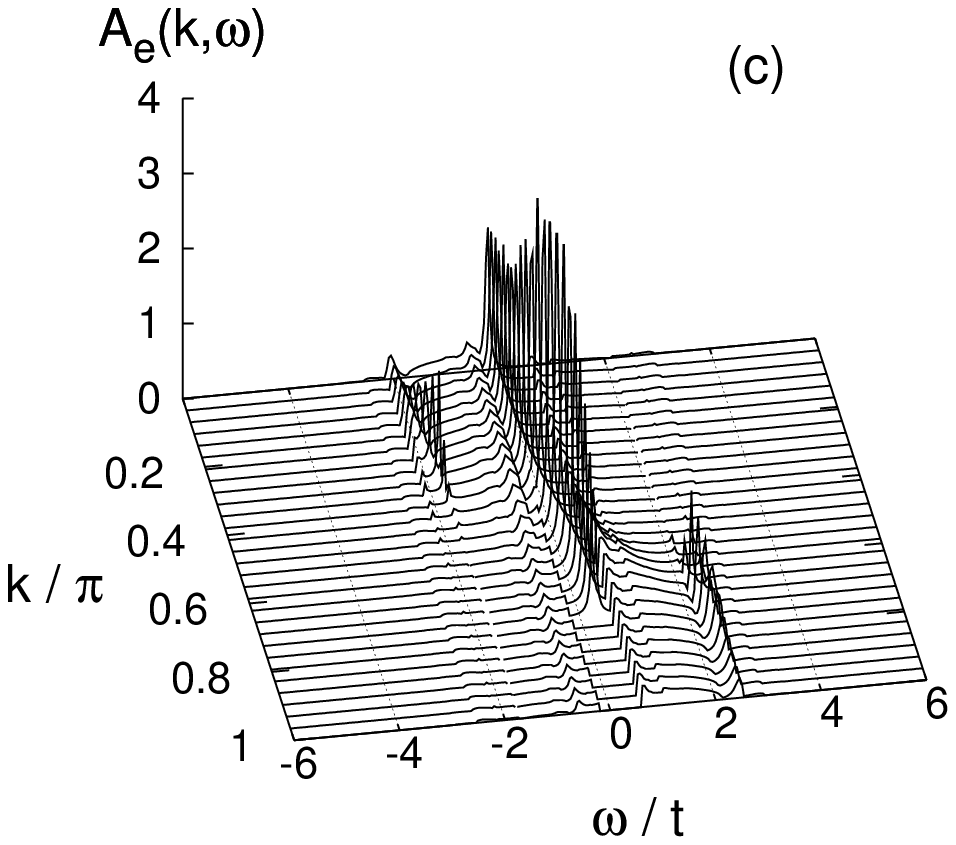}
  \includegraphics[width=0.495\textwidth]{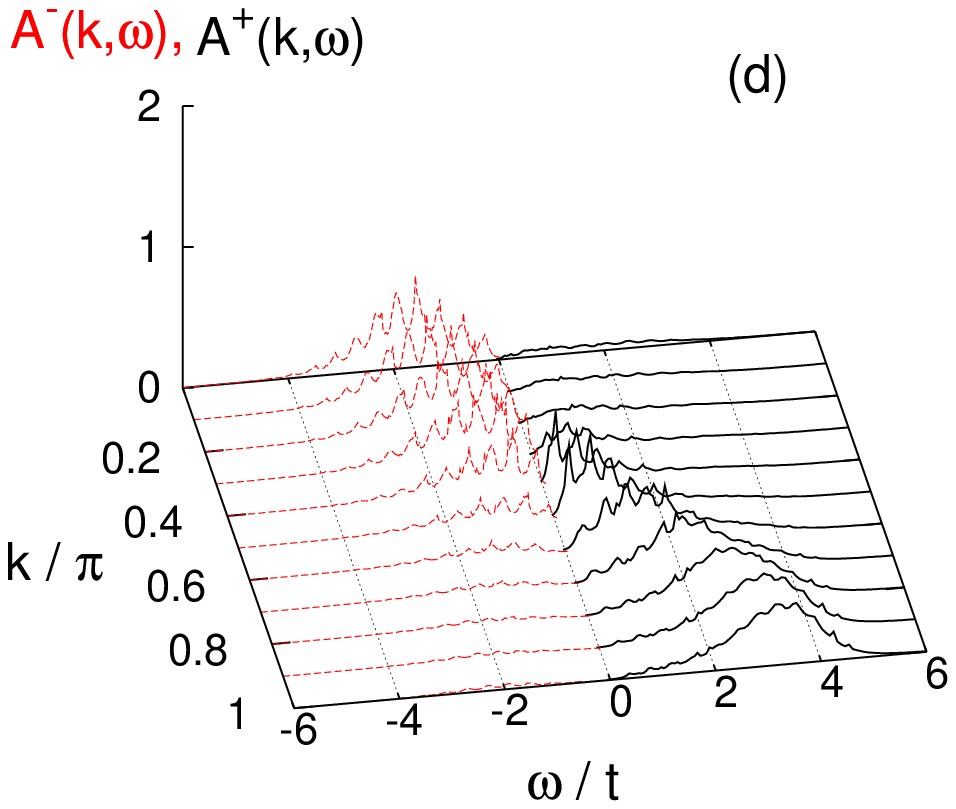}
  \caption{\label{fig:ic_w0.4}%
    (colour online)
    Spectral functions from the variational approach for different values of
    $\Ep$, $n=0.4$ and $\om_0/t=0.4$. The parameters are (a) $\Ep/t=0.1$,
    $R=1.5$, (b) $\Ep/t=4$, $R=0.1$ and (c) $\Ep/t=2$, $R=1.3$. The
    values of $R$ have been chosen according to
    figure~\ref{fig:free_energy_ic}(a). Panel~(d) shows cluster perturbation theory
    results for photoemission [$A^-(k,\om)$, dashed lines] and inverse photoemission
    [$A^+(k,\om)$] for $n=0.4$ and $\Ep=2$ (taken from \cite{HoWeAlFe05}).} 
\end{figure}
\begin{figure}
  \includegraphics[width=0.495\textwidth]{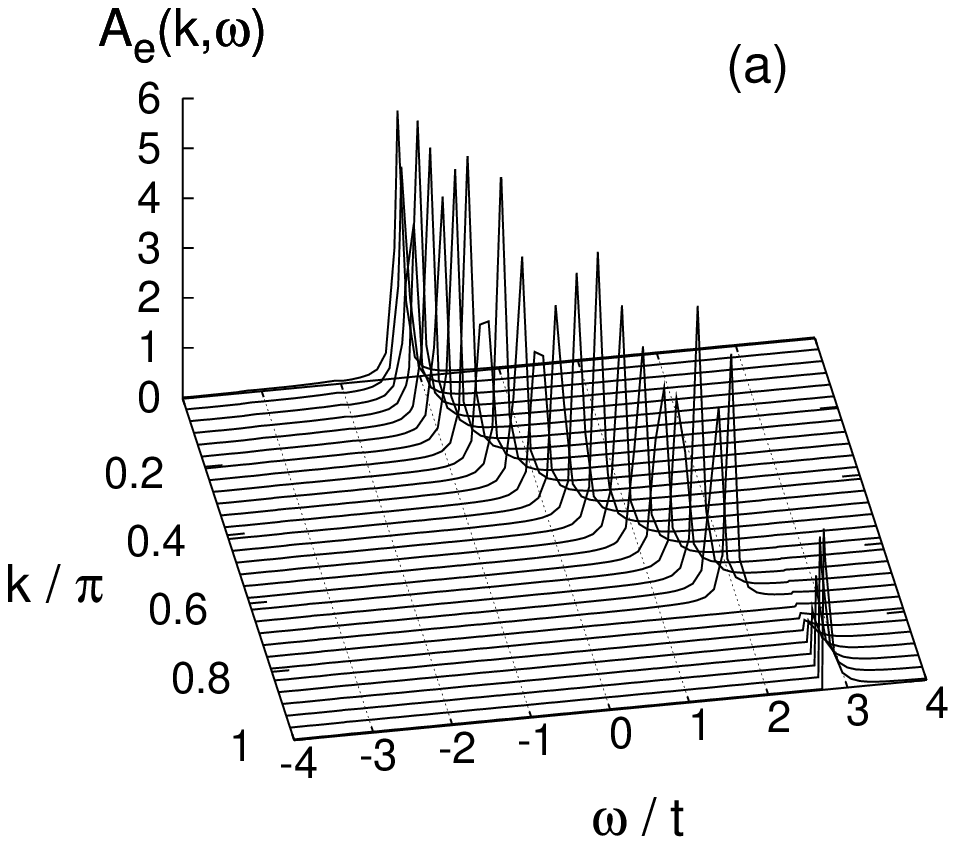}
  \includegraphics[width=0.495\textwidth]{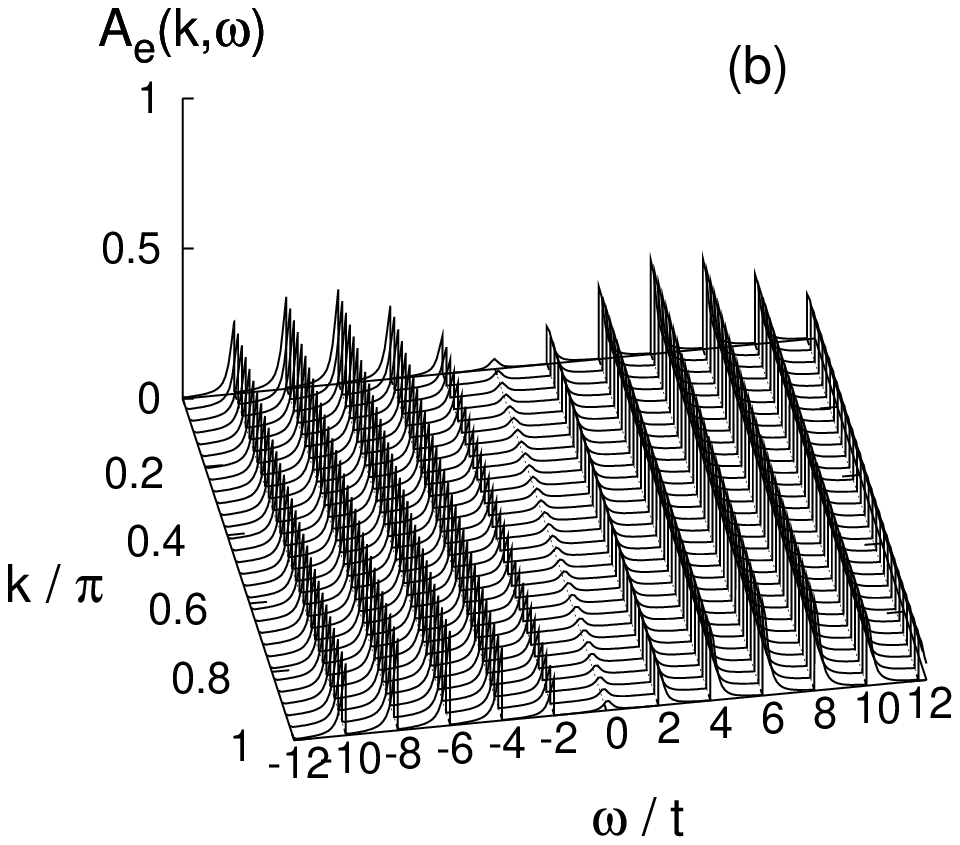}\\
  \includegraphics[width=0.495\textwidth]{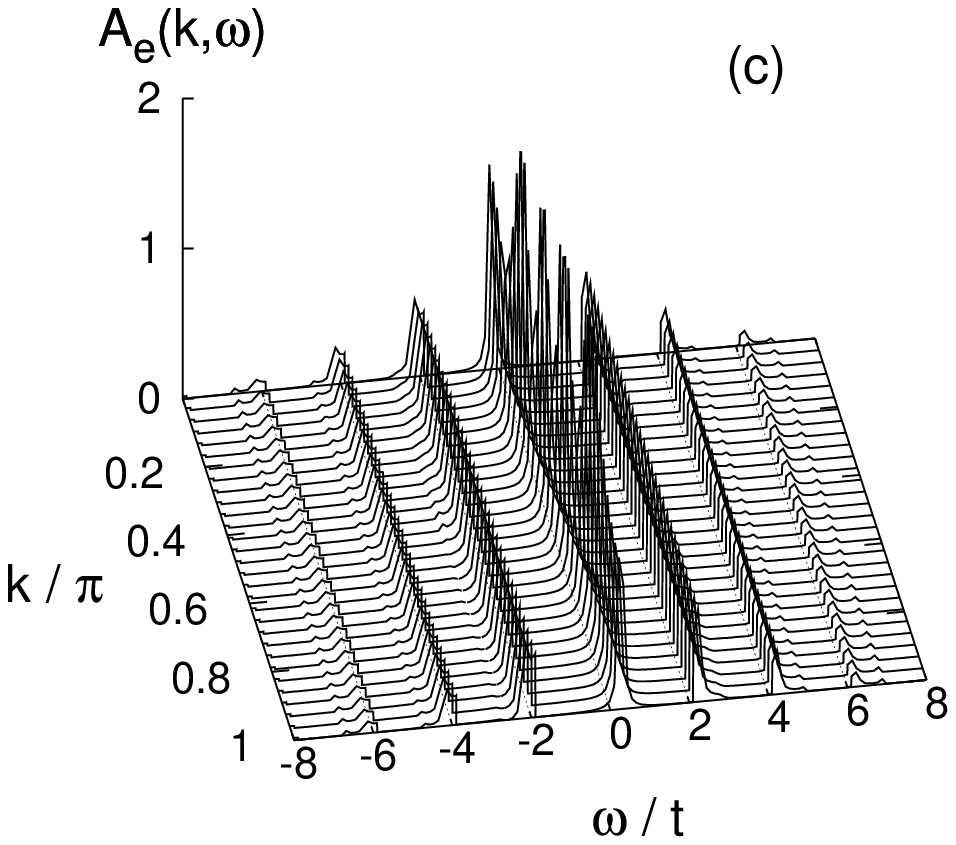}
  \includegraphics[width=0.495\textwidth]{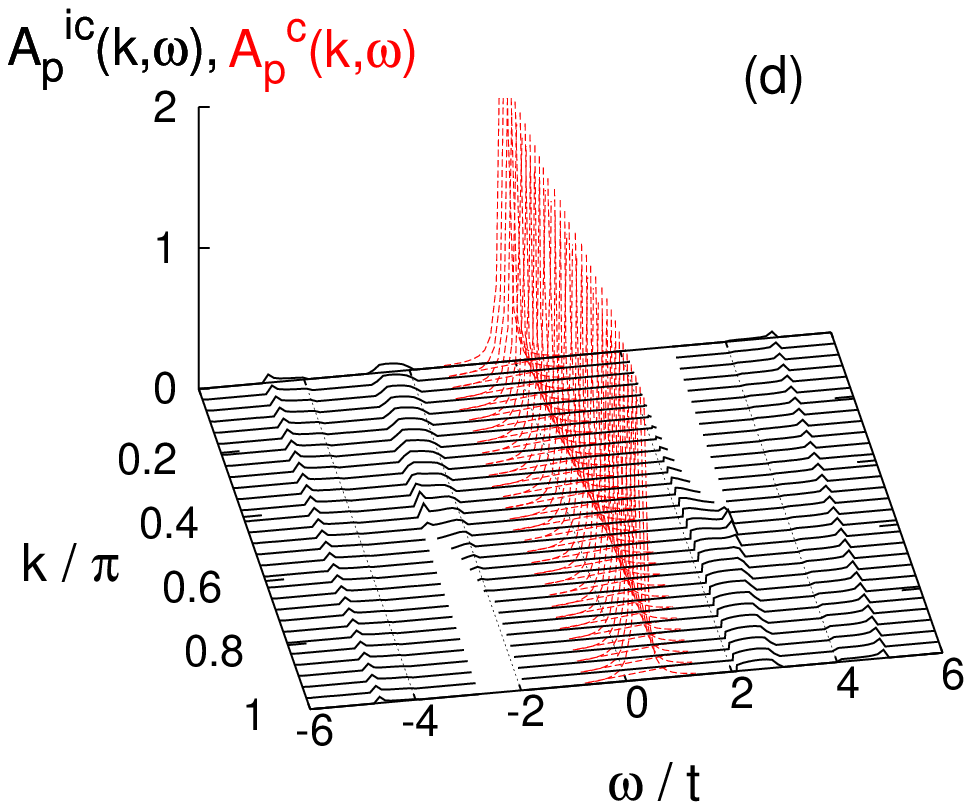}
  \caption{\label{fig:ic_w2.0}%
    (colour online)
    Spectral functions from the variational approach for different values of
    $\Ep$, $n=0.4$ and $\om_0/t=2$. The parameters are (a) $\Ep/t=0.1$,
    $R=0.7$, (b) $\Ep/t=8$, $R=0.2$, (c) and (d) $\Ep/t=4$, $R=0.5$. The
    values of $R$ have been chosen according to
    figure~\ref{fig:free_energy_ic}(b). In the white regions in (d)
    $\Ap^\mathrm{ic}<0$ (see text).  }
\end{figure}

As seen in the preceding sections, both the WC and SC approximations are not capable of
accounting for the recently discussed carrier density-driven cross-over from
a polaronic system to a metallic system with phonon-dressed electrons
\cite{HoNevdLWeLoFe04,HoWeAlFe05}. In fact, the electronic spectrum always
remains WC/SC-like in character. In order to obtain a reasonable, analytical
description of the IC regime, we therefore use the variational approach
proposed in section~\ref{sec:interm-coupl}.

Figure~\ref{fig:free_energy_ic} shows results for the total energy per site
as a function of the variational parameter $R$ for different values of $\Ep$.
The evolution is very similar to the large-to-small polaron cross-over in the
one-electron case \cite{FeLoWe97}.  Note that the present results have been
obtained using the Hartree approximation, as discussed in
section~\ref{sec:interm-coupl}.

For weak coupling $\Ep$, in the adiabatic case
[figure~\ref{fig:free_energy_ic}(a)], we find a minimum in the total energy at
a finite value of $R$. Upon increasing $\Ep$, a second local minimum starts
to develop near $R=0$, associated with the small-polaron state which becomes
the ground state in the SC limit. The jump of the optimal value of $R$ from a
large to a small value at a critical $\Ep$ suggests a first-order transition from an
extended to a small-polaron state. However, such a sharp transition is absent in
the single-polaron case, and not expected for $n>0$ either. The discontinuous
cross-over appearing in our results for $\om_0/t\lesssim1$ is a consequence of
the approximation used.

In contrast, in the non-adiabatic regime [figure~\ref{fig:free_energy_ic}(b)],
there exists only a single minimum, which shifts to smaller $R$ with
increasing coupling $\Ep$. Moreover, compared to
figure~\ref{fig:free_energy_ic}(a), the dependence of the total energy on $R$
is much weaker.

Apart from the comparison of the spectral function with other methods
presented below, variational approaches---often not capable of yielding
dynamic properties---are usually judged by the total energy as opposed to
exact data. Figure~\ref{fig:E0_Ep}(a) shows the second-order results for the
total energy as a function of $\Ep$, using the optimal values of the
parameter $R$ as determined from the Hartree approximation. Clearly, the
agreement between our IC approach and results from quantum Monte Carlo
\cite{HoNevdLWeLoFe04} is very good at weak and strong coupling (the
variational approach reproduces the WC and SC limits of
sections~\ref{sec:weak-coupling} and~\ref{sec:strong-coupling}), whereas
there are notable deviations at IC. The missing higher-order
corrections---causing the violation of the sum rule discussed earlier---lead
to generally overestimated values of the energy. Note that the agreement of
the energy with exact results is even better in the non-adiabatic regime
(not shown).

To illustrate the density-driven cross-over from (large) polarons to slightly
dressed electrons (scattered by diffusive phonons), figure~\ref{fig:E0_Ep}(b)
reports exact numerical results
for the renormalized kinetic energy. The latter may serve as a measure for the
carrier mobility. The increase as a function of $n$ may be interpreted as
originating from the overlap of the displacement clouds surrounding the
carriers and, finally, the dissociation of the polaronic
quasiparticles at large $n$.

We again begin the discussion of the spectral functions with the adiabatic
case $\om_0/t=0.4$. Following the discussion of section~\ref{sec:interm-coupl},
we determine the optimal $R$ from the position of the minimum of the total
energy for given $\Ep$ and $n$ (figure~\ref{fig:free_energy_ic}).

Figures~\ref{fig:ic_w0.4}(a) and (b) resemble closely to the WC and SC results
of figures~\ref{fig:wc_density_w0.4}(d) and \ref{fig:sc_density_w0.4}(b),
respectively. However, for IC $\Ep/t=2$ [figure~\ref{fig:ic_w0.4}(c)], we find
a rather metallic spectrum with a broad main band crossing the Fermi level,
and with low-energy excitations available. The corresponding polaronic
spectrum (not shown) reveals that the
variational approach correctly predicts the absence of well-defined polaronic
quasiparticles, as suggested by the non-negligible incoherent contributions
in $\Ap(k,\om)$, lying close in energy to the coherent band. This is in
contrast to the SC case, where polaronic quasiparticles dominate, and in
which the coherent band is well separated from the incoherent excitations. Note
that in figure~\ref{fig:ic_w0.4}(c) [and also in
figures~\ref{fig:ic_w2.0}(c) and~(d)] the incoherent part becomes
slightly negative for $\om\gtrsim -W$ ($\om\lesssim W$) at large (small) $k$,
which is an artifact of our approximation. Remarkably, the overall features
of the spectrum in figure~\ref{fig:ic_w0.4}(c) are very similar to the
corresponding numerical results in \cite{HoWeAlFe05}, reproduced in
figure~\ref{fig:ic_w0.4}(d).

Finally, in the non-adiabatic regime, it has been found numerically that small
polarons remain the correct quasiparticles even at large band fillings $n$
\cite{CaGrSt99,HoNevdLWeLoFe04}. Again, the variational approach is able to
describe the physics correctly. In particular, figures~\ref{fig:ic_w2.0}(c)
and~(d) reveal that a dominant coherent band, well separated from incoherent
excitations and having a relatively small bandwidth, exists even at IC.

As for the WC and SC cases discussed above, we have also checked the total
spectral weight in the present case. We find that the deviations from the
exact value of unity are largest for WC, whereas the sum rule is fulfilled
to within 1-10\% (depending on $\om_0$ and $\Ep$) at IC and SC.

\section{Conclusion}\label{sec:conclusion}

We have presented an analytical treatment of the one-dimensional spinless
Holstein model based on calculations of the self-energy in the framework of
the generalized Matsubara functions. To connect the analytical results to
previous numerical ones, the electronic spectral function determining the
photoemission spectrum has been computed for finite carrier concentrations in
dependence on the electron-phonon coupling strength and the phonon frequency.

In the strong-coupling limit, the electron spectral function has been
deduced from the spectral function of small polarons. However, it was shown
that the electron picture and the small-polaron picture both cease to be
correct if we approach the intermediate-coupling regime from
the weak-- and strong-coupling side, respectively.

To describe the cross-over from the strong-- to the weak-coupling limit, an
interpolation scheme based on a generalization of the Lang-Firsov canonical
transformation has been proposed. The latter was defined for each set of
model parameters by a distance $R$ which characterizes the charge
distribution across the polaron volume. The parameter $R$, as deduced from
the minimum of the total energy in the first approximation, was shown to increase
with decreasing $\Ep$---corresponding to a cross-over from large to small
polarons---and, at the same time, the coupling dependence of $R$ was found to
be stronger in the adiabatic than in the non-adiabatic case.

The spectral functions calculated at weak, strong and intermediate coupling
are in a good agreement with recent numerical calculations.
Moreover, analytical results deduced by means of the self-energy
calculations enabled us to distinguish between the coherent and incoherent
parts of the spectrum. Most importantly, starting from the strong-coupling
limit, it was shown that the spectral weight of the incoherent polaron
spectrum increases with decreasing coupling $\Ep$, and that the energy
separation of the incoherent peaks from the coherent spectrum is continuously
reduced in the adiabatic regime. On the contrary, the coherent part of the
electronic (photoemission) spectra is reduced by a factor $\rme^{-g^2}$, and
the incoherent part representing the phonon-assisted photoemission processes
becomes increasingly dominant with increasing coupling. 

Finally, at intermediate
coupling and finite carrier densities, our results support recent numerical
findings which suggest that the system can no longer be described in terms of
(small) polaronic quasiparticles \cite{HoNevdLWeLoFe04,HoWeAlFe05}.


\ack

We want to thank F.~X.~Bronold, H.~Sormann and G.~Wellein for helpful
discussions, and A.~S.~Alexandrov for valuable comments on the manuscript.
Furthermore, we would like to thank G.~Hager for providing us with the DMRG
data shown in figure~\ref{fig:E0_Ep}(b). 
 This work was supported by the Austrian Science fund
project P15834, the DFG through SPP1073, the DFG and the Academy of Sciences
of the Czech Republic (ASCR) under Grant Nr.~436 TSE 113/33/0-2, and the
Bavarian KONWIHR.  One of us (MH) is grateful to HPC-Europa. MH and HF
acknowledge the hospitality of the Institute of Physics, ASCR, Prague.



\section*{References}

\begin{thebibliography}{10}

\bibitem{mx}
A. R. Bishop and B. I. Swanson, Los Alamos Science {\bf 21} 133 (1993).

\bibitem{org}
I. H. Campbell and D. L. Smith, Solid State Phys. {\bf 55}, 1 (2001). 

\bibitem{htc}
A. S. Alexandrov and N. F. Mott, {\it Polarons and Bipolarons}
(World Scientific, Singapore, 1995).

\bibitem{mang} 
G. Zhao, K. Conder, H. Keller, and K. A. M\"{u}ller,
Nature (London) {\bf 381}, 676 (1996).

\bibitem{edwards}
D. M. Edwards, Adv. Phys. {\bf 51}, 1259 (2002).

\bibitem{Ho57}
T. Holstein, Ann. Phys. (N.Y.) {\bf 8}, 325 (1959).

\bibitem{MeScGu94}
V. Meden, K. Sch{\"o}nhammer, and O. Gunnarsson, Phys. Rev. B {\bf 50},  11 179
(1994).

\bibitem{Su72}
  H. Sumi, J. Phys. Soc. Jpn. {\bf 33},  327  (1972); 
\item[]
  S. Ciuchi, F. de~Pasquale, S. Fratini, and D. Feinberg, Phys. Rev. B {\bf 56},
  4494  (1997).

\bibitem{RT92}
J. Ranninger and U. Thibblin, Phys. Rev. B {\bf 45}, 7730 (1992).

\bibitem{WRF96} 
G. Wellein, H. R\"oder, and H. Fehske, Phys. Rev. B {\bf 53}, 9666 (1996).

\bibitem{KTB02}
L.-C. Ku, S. A. Trugman, and J. Bon\v{c}a, Phys. Rev. B {\bf 65}, 174306 
(2002).

\bibitem{HoEvvdL03}
M. Hohenadler, H.~G. Evertz, and W. {von der Linden}, Phys. Rev. B {\bf 69},
024301 (2004);
\item[]
M. Hohenadler, M. Aichhorn, and W. {von der Linden}, Phys. Rev. B {\bf 71},
014302 (2005).

\bibitem{Spencer}
P.~E. Spencer, J.~H. Samson, P.~E. Kornilovitch, and A.~S. Alexandrov,
Phys. Rev. B {\bf 71}, 184310 (2005).
  
\bibitem{BuMKHa98}
R.~J. Bursill, R.~H. McKenzie, and C.~J. Hamer, Phys. Rev. Lett. {\bf 80},
5607  (1998);
\item[] 
H. Fehske, M. Holicki, and A. Wei{\ss}e, Adv. Sol. State Phys. {\bf 40},  235
(2000);
\item[]
M. Capone and S. Ciuchi, Phys. Rev. Lett {\bf 91}, 186405 (2003); 
\item[]
S. Sykora {\it et al.}, Phys. Rev. B {\bf 71},  045112  (2005).

\bibitem{HuZheng}
  Q. S. Hu and H. Zheng, Eur. Phys. J. B {\bf 28}, 255 (2002); 
\item[]
  H. Zheng, J. Phys.: Condens. Matter {\bf 36}, 9405 (2003).

\bibitem{Datta}
  S. Datta, A. Das, and S. Yarlagadda, Phys. Rev. B {\bf 71}, 235118 (2005).

\bibitem{CaGrSt99}
M. Capone, M. Grilli, and W. Stephan, Eur. Phys. J. B {\bf 11},  551  (1999).

\bibitem{HoNevdLWeLoFe04}
M. Hohenadler {\it et~al.}, Phys. Rev. B {\bf 71},  245111  (2005).

\bibitem{HoWeAlFe05}
M. Hohenadler, G. Wellein, A. Alvermann, and H. Fehske, cond-mat/0505559,
Physica B (2006); 
\item[]
G. Wellein {\it et~al.}, 
cond-mat/0505664, Physica B (2006).

\bibitem{LangFirsov}
I.~G. Lang and Y.~A. Firsov, Zh. Eksp. Teor. Fiz. {\bf 43},  1843  (1962),
[Sov. Phys. JETP {\bf 16}, 1301 (1962)].

\bibitem{KaBa62}
L.~P. Kadanoff and G. Baym, {\em Quantum Statistical Mechanics}
(Benjamin-Cumming, Reading, MA, 1962).

\bibitem{BBTy62}
V.~L. {Bonch-Bruevich} and S.~V. Tyablikov, {\em The Green Function Method in
Statistical Mechanics} (North-Holland Publ. Co., Amsterdam, 1962).

\bibitem{Sc66}
J. Schnakenberg, Z. Phys. {\bf 190},  209  (1966).

\bibitem{Lo94}
J. Loos, Z. Phys. B {\bf 96},  149  (1994).

\bibitem{FeLoWe97}
H. Fehske, J. Loos, and G. Wellein, Z. Phys. B {\bf 104},  619  (1997); 
Phys. Rev. B {\bf 61},  8016  (2000).

\bibitem{Ma90}
G.~D. Mahan, {\em Many-Particle Physics}, 2nd  ed. (Plenum Press, New York,
1990).

\bibitem{AlRa92}
A.~S. Alexandrov and J. Ranninger, Phys. Rev. B {\bf 45},  13 109  (1992).

\bibitem{Al_book}
A.~S. Alexandrov, {\em Theory of superconductivity: from Weak to Strong
Coupling} (IoP Publishig, Bristol, 2003), pp.110-116.

\bibitem{Zheng}
H. Zheng, D. Feinberg, and M. Avignon, Phys. Rev. B {\bf 51}, 11 557 (1990).

\bibitem{St96}
W. Stephan, Phys. Rev. B {\bf 54}, 8981 (1996);
\item[]
G. Wellein and H. Fehske, Phys. Rev. B {\bf 55}, 4513 (1997).

\bibitem{Korni}
P.~E. Kornilovitch, Europhys. Lett. {\bf 59}, 735 (2002).

\end{thebibliography}

\end{document}